\documentclass[prd,aps,
preprint,
tightenlines,
superscriptaddress,showpacs,nofootinbib,
eqsecnum,amsfonts,amsmath,epsf]{revtex4}

 \newcommand{\bs}{\begin{subequations}}
\newcommand{\es}{\end{subequations}} \newcommand{\beq}{\begin{equation}}
\newcommand{\eeq}{\end{equation}}
\newcommand{\bes}{\begin{subequations}}
\newcommand{\ees}{\end{subequations}}
\newcommand{\bea}{\begin{align}} \newcommand{\eea}{\end{align}}
\newcommand{\ba}{\begin{array}} \newcommand{\ea}{\end{array}}
\newcommand{\beqn}{\begin{align*}}
\newcommand{\eeqn}{\end{align*}}

\usepackage{bm,graphicx,color,epsfig}
\usepackage{subfigure}
\usepackage{amsfonts}
\usepackage{amsopn}
\usepackage{amssymb}
\usepackage{pxfonts}
\usepackage{rotating} 
\usepackage{setspace}

\usepackage{amssymb}
\usepackage{lscape}
\usepackage{bm}
\usepackage[varg]{txfonts}

\allowdisplaybreaks

\begin{document}
\title{Inspiralling compact binaries in quasi-elliptical orbits:\\ The
complete third post-Newtonian energy flux } \date{\today} 
\author{K G Arun}
\email{arun@iap.fr} \affiliation{Raman Research Institute, Bangalore
560 080, India} 
\affiliation{${\mathcal{G}}{\mathbb{R}}
\varepsilon{\mathbb{C}}{\mathcal{O}}$, Institut d'Astrophysique de Paris
--- C.N.R.S., 98$^{\text{bis}}$ boulevard Arago, 75014 Paris, France}
\affiliation{LAL, Universit{\'e} Paris-Sud, IN2P3/CNRS, Orsay, France}
\author{Luc Blanchet} \email{blanchet@iap.fr}
\affiliation{${\mathcal{G}}{\mathbb{R}}
\varepsilon{\mathbb{C}}{\mathcal{O}}$, Institut d'Astrophysique de Paris
--- C.N.R.S., 98$^{\text{bis}}$ boulevard Arago, 75014 Paris, France}
\author{Bala R Iyer} \email{bri@rri.res.in} \affiliation{Raman Research
Institute, Bangalore 560 080, India} \author{Moh'd S S
Qusailah}\email{mssq@rri.res.in}
 \affiliation{Raman Research Institute, Bangalore
560 080, India}
\affiliation{University of Sana, Yemen}
\begin{abstract}
\end{abstract}
\preprint{gr-qc/yymmnnn}

\narrowtext
\begin{abstract}
The instantaneous contributions to the third post-Newtonian (3PN)
 gravitational wave
luminosity from the inspiral phase of a binary system of compact
objects moving in a quasi elliptical orbit is computed using the
multipolar post-Minkowskian wave generation formalism. The necessary
inputs for this calculation include the 3PN accurate mass quadrupole
moment for general orbits and the mass octupole and current quadrupole
moments at 2PN. Using the recently obtained 3PN quasi-Keplerian
representation of elliptical orbits the flux is averaged over the
binary's orbit. Supplementing this by the important hereditary
contributions arising from tails, tails-of-tails and tails squared
terms calculated in a previous paper, the complete 3PN energy flux is
obtained. The final result presented in this paper would be needed for
the construction of ready-to-use templates for binaries moving on
non-circular orbits, a plausible class of sources not only for the
space based detectors like LISA but also for the ground based ones.
\end{abstract}

\maketitle

\noindent

\section{Introduction}\label{intro}
%
Inspiralling compact binaries, one of the prototype sources for laser
interferometric gravitational wave (GW) detectors, are usually
modelled as moving in quasi-circular orbits. This is justified since
gravitational radiation reaction, under which it inspirals,
circularizes the orbit towards the late stages of
inspiral~\cite{PM63,Pe64}. This late phase of inspiral and the ensuing
merger phase offer promises for the GW interferometric detectors. The
recently discovered double pulsar system~\cite{Burgay03,kalogera} has
an eccentricity as low as $0.088$ consistent with the circular orbit
assumption for the late inspiral and pre-merger phases, believed to be
reasonable enough for most of the binary systems made of neutron stars
or black holes (BHs).

The theoretical modelling of the binary's phase evolution to a very
high precision is called the phasing formula. This is the basic
theoretical ingredient used in the construction of search templates
for GW using matched filtering~\cite{Wainstein}. The two key inputs
required for the construction of templates for binaries moving in
quasi-circular orbits in the adiabatic approximation are the orbital
energy and the GW luminosity (energy flux). These are computed using a
cocktail of approximation schemes in general relativity. The schemes
include the multipole decomposition, the post-Minkowskian expansion of
the gravitational field or non-linearity expansion in Newton's
constant $G$, the post-Newtonian expansion in $v/c$, and the far-zone
expansion in powers of $1/R$, where $R$ is the distance from the
source (see~\cite{Bliving} for a recent review).

Though the garden variety binary sources of GWs for terrestrial
 laser interferometric GW detectors are those moving in
quasi-circular orbits, there is an increased recent interest in
inspiralling binaries moving in {\it quasi-eccentric} orbits.
Astrophysical scenarios currently exist which lead to binaries with
non-zero eccentricity
in the GW detector bandwidth, both terrestrial and space-based. For
instance, inner binaries of hierarchical triplets undergoing Kozai
oscillations~\cite{KozaiOsc62} could not only merge due to
gravitational radiation reaction but a good fraction
($\sim 30$\%) of them will have eccentricity greater than about $0.1$
as they enter the sensitivity band of advanced ground based
interferometers~\cite{Wen02}. Almost all the above systems possess
eccentricities below $0.2$ at $40$ Hz and below $0.02$ at $200$
Hz. The population of stellar mass binaries in globular clusters is
expected to have a thermal distribution of
eccentricities~\cite{Benacquista02}. In a study on the growth of
intermediate BHs~\cite{GMHimbh04} in globular clusters it was found
that the binaries have eccentricities between $0.1$ and $0.2$ in the
LISA bandwidth. Though, supermassive black hole binaries are powerful
GW sources for LISA, it is not yet conclusive if they would be in
quasi-circular or quasi-eccentric orbits~\cite{ThBrag76}. If a Kozai
mechanism is at work, these supermassive BH binaries could be in
highly eccentric orbits and merge within the Hubble
time~\cite{BLSKozai02}. Sources of the kind discussed above provide
the prime motivation to investigate higher post-Newtonian order
modelling for quasi-eccentric binaries.

The GW energy flux or luminosity from a
system of two point masses in elliptic motion was first computed by
Peters and Mathews at Newtonian order~\cite{PM63,Pe64}. The post-Newtonian (PN)
corrections to the gravitational wave flux at 1PN and 1.5PN
  were provided
in~\cite{WagW76,BS89,JunkS92,BS93,RS97} and used to study the
associated evolution of orbital elements using the 1PN
``quasi-Keplerian'' representation of the binary's orbit 
~\cite{DD85}. Gopakumar and Iyer~\cite{GI97,GI02}
further extended these results to 2PN order using the generalized
quasi-Keplerian representation developed in
Ref.~\cite{DS88,SW93,Wex95}. The results for the energy flux and
waveform presented in~\cite{GI97} was in perfect agreement with those
obtained by Will and Wiseman using a different
formalism~\cite{WWi96}. Recently, Damour, Gopakumar and
Iyer~\cite{DGI04} discussed an analytic method for constructing high
accuracy templates for the GW signals from the inspiral phase of
compact binaries moving on quasi-elliptical orbits.  They used an
improved ``method of variation of constants'' to combine the three
time scales involved in the elliptical orbit case, namely, orbital
period, periastron precession and radiation reaction time scales,
without making the usual approximation of treating the radiative time
scale as an adiabatic process.

The generation problem for gravitational waves at any PN
order requires the solution to two independent problems. The first
relates to the equation of motion of the binary and the second to the
far zone fluxes of energy, angular momentum and linear momentum. The
latter requires the computation of the relativistic mass and current
multipole moments to appropriate PN orders. The 3PN
equations of motion (EOM) required to handle gravitational wave
phasing turned out to be technically very involved due to the issues
related to the self-field regularization using Riesz or Hadamard
regularizations~\cite{JS99,BFeom}. Only by a deeper understanding of
the origin of the ambiguities in Hadamard regularization, and the use
of dimensional regularization
has the problem been uniquely
resolved~\cite{DJSdim,BDE04} and provided the value of the ambiguity
parameter $\omega_s$~\cite{JS99} or equivalently
$\lambda$~\cite{BFeom}. We thus have in hand the requisite 3PN EOM for
compact binaries moving in general orbits. The computation of the GW
luminosity at 3PN or $(v/c)^6$ beyond the leading Einstein quadrupole
formula crucially requires the computation of the 3PN accurate mass
quadrupole moment. For its completion the same technique as in the EOM was
successfully applied, namely, to compute using Hadamard's
regularization all the terms except a few terms parametrized
by ambiguity parameters (which
turn out to be three, denoted $\xi$, $\kappa$ and
$\zeta$)~\cite{BIJ02,BI04}, and then to determine the value of these
parameters by computing the difference between the dimensional and
Hadamard regularizations~\cite{BI04,BDEI04,BDI04zeta,BDEI05}. These works
thus provide the fully determined 3PN accurate mass quadrupole for
general orbits -- the other important ingredient to compute the 3PN
accurate energy and angular momentum fluxes for inspiralling compact
binaries moving in general non-circular orbits. The 3.5PN phasing of
inspiralling compact binaries moving in quasi-circular orbits is now
complete and available for use in GW data
analysis~\cite{BFIJ02,BDEI04}. Note that the 3PN contribution to the
energy flux comes not only from the ``instantaneous'' terms discussed
in this paper but also includes ``hereditary'' contributions arising
from tails, tails of tails and tail-squared terms. A semi-analytical scheme
is proposed and discussed in detail in a companion
paper~\cite{ABIQ07Tail}\footnote{Hereafter Ref.~\cite{ABIQ07Tail} will
be called Paper~I.} to evaluate these history-dependent contributions.

In this paper, for binaries moving in elliptical orbits, we compute
all the instantaneous contributions to the 3PN accurate GW energy
flux. The orbital average of this flux will be obtained using the
3PN quasi-Keplerian parametrization of the
binary's orbital motion recently constructed by Memmesheimer, Gopakumar and
Sch\"afer~\cite{MGS04}. We shall supplement these by contributions
from the hereditary terms computed in Paper~I. The final expression
will represent gravitational waves from a binary evolving negligibly
under gravitational radiation reaction, including precisely up to 3PN
order the effects of eccentricity and periastron precession during
epochs of inspiral when the orbital parameters are essentially
constant over a few orbital revolutions. It also represents the first
step towards the discussion of the {\it quasi-elliptical} case: the
evolution of the binary in an elliptical orbit under gravitational
radiation reaction. The present work extends the circular orbit
results at 2.5PN~\cite{B96} and 3PN~\cite{BIJ02,BDEI04} to the elliptical
orbit case. Further, it extends earlier works on instantaneous
contributions for binaries moving in elliptical orbits at
1PN~\cite{BS89,JunkS92} and 2PN~\cite{GI97} to 3PN order. Similarly, Paper~I
extends hereditary contributions at
1.5PN~\cite{BS93}
to 2.5PN order and 3PN, where
the 3PN hereditary contributions comprise the tails of tails and are
extensions of Refs.~\cite{B98quad,B98tail} for circular orbits to the
elliptical orbit case.

In Sect.~\ref{fz} we begin with the structure of the far-zone flux of
energy, use expressions relating the radiative moments to the source
moments and decompose the energy flux expression into its
instantaneous and hereditary parts. Section~\ref{input} lists all the
requisite multipole moments in standard harmonic coordinates for
binaries moving in general (non-circular) orbits. Section~\ref{EqnOM}
introduces the 3PN equations of motion which are necessary to
handle the time derivatives of the moments. Section~\ref{EF-Sharm}
discusses the computation of the instantaneous terms in the energy
flux and Sect.~\ref{EF-alt} recasts the flux in modified harmonic
coordinates (without logarithms at 3PN order) and Arnowitt, Deser and Misner (ADM) coordinates. Section~\ref{QKrepr} summarises the 3PN quasi-Keplerian
representation required to average the flux expression over an
orbit. Section~\ref{avgE} exhibits the orbital average of the energy
flux in modified harmonic coordinates and ADM coordinates, and finally
provides an expression of the complete energy flux
 in terms
of gauge invariant variables.
%
\section{The far zone flux of energy}\label{fz}
%
In this section, we discuss the computation of the 3PN accurate energy
flux for general isolated sources. Starting from the expression for
the far zone flux in terms of the radiative multipole moments and
using the relations connecting the radiative multipole moments to the
source moments, we write the resultant structure of the GW energy
flux.

Following Thorne~\cite{Th80}, the expression for the 3PN accurate far
zone energy flux $\mathcal{F}\equiv (d\mathcal{E}/d t)^\mathrm{GW}$ in
terms of symmetric trace-free (STF) radiative multipole moments reads
as\footnote{The shorthand $\mathcal{O}(n)$ is used throughout and
indicates that the post-Newtonian remainder is of order of
$\mathcal{O}(c^{-n})$.}
\begin{align}
\mathcal{F} &={G\over c^5}\Biggl\{ {1\over 5} U^{(1)}_{ij}
U^{(1)}_{ij}\nonumber\\ &\quad+{1\over c^2} \left[ {1\over 189}
U^{(1)}_{ijk} U^{(1)}_{ijk} +{16\over 45} V^{(1)}_{ij}
V^{(1)}_{ij}\right]+{1\over c^4} \left[ {1\over 9072} U^{(1)}_{ijkm}
U^{(1)}_{ijkm}+{1\over 84} V^{(1)}_{ijk}
V^{(1)}_{ijk}\right]\nonumber\\ &\quad+{1\over c^6}\left[ {1\over
594000} U^{(1)}_{ijkmn} U^{(1)}_{ijkmn}+{4\over 14175} V^{(1)}_{ijkm}
V^{(1)}_{ijkm}\right]+\mathcal{O}(8)\Biggr\}\,.\label{EFUV}
\end{align}
In the above $U_{L}$ and $V_{L}$ (where $L=i_1i_2\cdots i_l$
represents a multi-index composed of $l$ spatial indices) are the mass-type
and current-type radiative multipole moments respectively and
$U_{L}^{(l)}$ and $V_{L}^{(l)}$ denote their $l^\mathrm{th}$ time
derivatives. The moments are functions of retarded time $U\equiv
T-R/c$ in radiative coordinates. 

In the multipolar-post-Minkowskian (MPM) formalism, the radiative
moments $U_{L}$ and $V_{L}$ can be re-expressed in terms of the source
moments to an accuracy sufficient for the computation of the energy
flux. For the flux to be complete up to 3PN approximation, one must
compute the mass type radiative quadrupole $U_{ij}$ to 3PN accuracy,
mass octupole $U_{ijk}$ and current quadrupole $V_{ij}$ to 2PN
accuracy, mass hexadecapole $U_{ijkm}$ and current octupole $V_{ijk}$
to 1PN accuracy and finally $U_{ijkmn}$ and $V_{ijkm}$ to Newtonian
accuracy.

The relations connecting the different radiative moments $U_{L}$ and
$V_{L}$ to the corresponding source moments $I_L$ and $J_L$ are given
below. For the 3PN mass quadrupole moment we
have~\cite{BD92,B96,B98quad,B98tail}
\begin{align}
U_{ij}(U) &= I^{(2)}_{ij} (U) + {2GM\over c^3} \int_{0}^{+\infty} d
\tau \left[ \ln \left({c\tau \over 2r_0}\right)+{11\over2} \right]
I^{(4)}_{ij} (U-\tau) \nonumber \\
&+\frac{G}{c^5}\left\{-\frac{2}{7}\int_{0}^{+\infty} d\tau
I^{(3)}_{a\langle i}(U-\tau)I^{(3)}_{j\rangle a}(U-\tau) \right.\nonumber \\
&\qquad~ + {1 \over7}I^{(5)}_{a\langle i}I_{j\rangle a} - {5 \over7}
I^{(4)}_{a\langle i}I^{(1)}_{j\rangle a} -{2 \over7} I^{(3)}_{a\langle i}I^{(2)}_{j\rangle a} +{1
\over3}\varepsilon_{ab\langle i}I^{(4)}_{j\rangle a}J_{b}\nonumber\\ &\qquad~\left.
+4\left[W^{(2)}I_{ij}-W^{(1)}I_{ij}^{(1)}\right]^{(2)} \right\}\nonumber \\
&+2\left(\frac{G M}{c^3}\right)^2\int_{0}^{+\infty}d\tau I_{ij}^{(5)}
\left(U-\tau\right)\left[\ln^2\left({c\tau \over
2r_0}\right)+{57\over70} \ln\left({c\tau \over
2r_0}\right)+{124627\over44100}\right] +\mathcal{O}(7),\label{Uij}
\end{align}
where the bracket $<>$ surrounding indices denotes the STF projection,
and $\varepsilon_{abi}$ is the usual Levi-Civita symbol such that
$\varepsilon_{123}=+1$. The $I_{L}$'s and $J_L$'s are the mass and
current-type source moments (and $I_{L}^{(p)}$, $J_{L}^{(p)}$ denote
their $p^\mathrm{th}$ time derivatives), and $W$ is the monopole
corresponding to the set of ``gauge'' moments $W_L$,
using the same definitions as in \cite{BIJ02}.
In the above formula,
$M$ (which is in factor of the
tail integral at 1.5PN order and the tail-of-tail integral at 3PN) is the total ADM mass of the source. The
non-linear memory integral at 2.5PN is a time anti-derivative and will
become instantaneous in the energy flux. The moments needed at 2PN
order include only the dominant tails and are
\begin{subequations}\label{U}
\begin{align}
U_{ijk} (U) &= I^{(3)}_{ijk} (U) + {2GM\over c^3} \int_{0}^{+\infty}
d\tau\left[ \ln \left({c\tau \over 2r_0}\right)+{97\over60} \right]
I^{(5)}_{ijk} (U-\tau)+\mathrm{\mathcal{O}(5)} \,,\label{Uijk} \\V_{ij} (U)
&= J^{(2)}_{ij} (U) + {2GM\over c^3} \int_{0}^{+\infty} d \tau \left[
\ln \left({c\tau \over 2r_0}\right)+{7\over6} \right] J^{(4)}_{ij}
(U-\tau) + \mathcal{O}(5) \,.\label{Vij}
\end{align}\end{subequations}
For all the other moments required in the computation we need only the
leading order accuracy in the relation between radiative and source
moments, so that
\begin{subequations}\label{UVl}
\begin{align}
U_L (U) &= I^{(l)}_L(U) + \mathcal{O}(3)\,,\\
V_L (U) &= J^{(l)}_L(U) + \mathcal{O}(3)\,.
\end{align}
\end{subequations}

The constant length $r_0$ scaling the logarithm is the one introduced
in the general MPM formalism and has been chosen here to match with
the choice made in the computation of tails-of-tails
in~\cite{B98tail}. It is a freely specifiable constant, entering the
relation between the retarded time $U=T-R/c$ in radiative coordinates
and the corresponding retarded time $t_{\rm H}-r_{\rm H}/c$ 
in harmonic coordinates
(where $r_{\rm H}$ is the distance of the source in harmonic
coordinates). More precisely we have
\begin{equation}
U = t_{\rm H}- \frac{r_{\rm H}}{c}
-\frac{2\,G\,M}{c^3}\ln\left(\frac{r_{\rm H}}{r_0}\right) + \mathcal{O}(5)\,.
\label{Tr}\end{equation}

From Eqs~\eqref{Uij}-\eqref{UVl}, it is clear that the radiative
moments have two distinct contributions. One part depends on the
moments only at the retarded time, $U=T-R/c$; this part is referred to
as the ``instantaneous contribution'', and forms the subject matter of
the present paper. The second part on the other hand depends on the
dynamics of the system in its entire past, \textit{i.e.} at any
$U-\tau < U$, and is referred to as the ``hereditary
contribution''. Equally important but requiring a different treatment,
the hereditary contribution is dealt with in Paper~I as mentioned
earlier. We are thus allowed to write down explicitly the different
kinds of contributions to the far zone energy flux up to 3PN. We have,
\begin{equation}
\mathcal{F}=\mathcal{F}_{\rm
inst}+\mathcal{F}_\mathrm{hered}\,,
\label{tot}
\end{equation}
where the instantaneous contribution of interest in this paper is
explicitly given by
\begin{align}
\mathcal{F}_\mathrm{inst}&={G\over c^5}\biggl\{ {1\over 5}
I^{(3)}_{ij} I^{(3)}_{ij}\nonumber\\ &+{1\over c^2} \left[ {1\over
189} I^{(4)}_{ijk} I^{(4)}_{ijk} +{16\over 45} J^{(3)}_{ij}
J^{(3)}_{ij}\right]+{1\over c^4} \left[ {1\over 9072} I^{(5)}_{ijkm}
I^{(5)}_{ijkm}+{1\over 84} J^{(4)}_{ijk}
J^{(4)}_{ijk}\right]\nonumber\\
&+\frac{G}{c^5}\left[\frac{8}{5}I_{ij}^{(3)}\left(I_{ij}W^{(5)}+2
I_{ij}^{(1)}W^{(4)}-2I_{ij}^{(3)}W^{(2)}-I_{ij}^{(4)}W^{(1)}\right)\right.\nonumber\\
&\qquad\left.+\frac{2}{5} I_{i j}^{(3)} \left(-\frac{4}{7}
I_{ai}^{(5)} I_{aj}^{(1)}-I_{ai}^{(4)} I_{aj}^{(2)}-\frac{2}{7}
I_{ai}^{(3)} I_{aj}^{(3)}+\frac{1}{7} I_{ai}^{(6)} I_{aj}
\right)+\frac{2}{15} \varepsilon _{a b i}I_{a j}^{(5)}
J_b\, I_{ij}^{(3)}\right]\nonumber\\ &+{1\over c^6}\left[ {1\over 594000}
I^{(6)}_{ijkmn} I^{(6)}_{ijkmn}+{4\over 14175} J^{(5)}_{ijkm}
J^{(5)}_{ijkm}\right]+\mathcal{O}(8)\biggr\}\,.\label{inst}
\end{align}
The hereditary contribution is given in Sect.~III A of Paper~I.
We recall that it
is decomposed as
\begin{equation}
\mathcal{F}_\mathrm{hered}=\mathcal{F}_{\rm
tail}+\mathcal{F}_\mathrm{tail(tail)}+\mathcal{F}_\mathrm{(tail)^2}\,.
\label{heredtot}
\end{equation}
The quadratic-order (proportional to $G^2$) tails are given by
\begin{align}\label{Ftail}
\mathcal{F}_\mathrm{tail} &= \frac{4 G^2\,M}{5c^8}\,I_{ij}^{(3)}(U)\,
\int_0^{+\infty}d\tau\,I^{(5)}_{ij}(U-\tau)\biggl[\ln\left(\frac{c\tau}{2r_0}\right)
+\frac{11}{12}\biggr]\nonumber\\ &+ \frac{4G^2\,M}{189
c^{10}}\,I_{ijk}^{(4)}(U)\,\int_0^{+\infty}d\tau\,I^{(6)}_{
ijk}(U-\tau)\biggl[\ln\left(\frac{c\tau}{2r_0}\right)+\frac{97}{60}\biggr]\nonumber\\
&+ \frac{64G^2\,M}{45
c^{10}}\,J_{ij}^{(3)}(U)\,\int_0^{+\infty}d\tau\,J^{(5)}_{
ij}(U-\tau)\biggl[\ln\left(\frac{
c\tau}{2r_0}\right)+\frac{7}{6}\biggr]\,,
\end{align}
and the cubic-order tails (proportional to $G^3$) read
\begin{subequations}
\label{Ftailtail}\begin{align}
\mathcal{F}_\mathrm{tail(tail)} &=
\frac{4G^3M^2}{5 c^{11}}\,I_{ij}^{(3)}(U)\,\int_0^{+\infty}d\tau\,I^{(6)}_{
ij}(U-\tau)\biggl[\ln^2\left(\frac{
c\tau}{2r_0}\right)+\frac{57}{70}\ln\left(\frac{
c\tau}{2r_0}\right)+\frac{124627}{44100}\biggr],\\
\mathcal{F}_\mathrm{(tail)^2} &=
\frac{4G^3M^2}{5 c^{11}}\left(\int_0^{+\infty}d\tau\,I^{(5)}_{
ij}(U-\tau)\biggl[\ln\left(\frac{
c\tau}{2r_0}\right)+\frac{11}{12}\biggr]\right)^2\,.
\end{align}\end{subequations}
All the tail contributions are thoroughly computed in Paper~I and we
shall use those results to obtain the complete GW energy flux in Sec.~\ref{avgE}.
%
\section{The multipole moments of compact binary systems}\label{input}
%
We provide, in this Section, the requisite multipole moments needed
for the computation of the 3PN accurate energy flux for compact
binaries in the \textit{standard harmonic} coordinate system. By
standard harmonic coordinates we refer to the specific coordinate
system which has been used consistently in previous
works~\cite{BFeom,BDE04,BIJ02,BDEI04,BI03CM,BDEI05,BFIJ02,BI04}. We recall that these coordinates
contain some {\it logarithms} at the 3PN level both in the equations
of motion of the binary~\cite{BFeom,BI03CM} and in their multipole
moments~\cite{BIJ02,BI04,BDEI04}. Later, we shall also define some \textit{modified
harmonic} coordinates which do not involve such logarithms at the 3PN
order.

The multipole moments are generalisations to non-circular orbits of
the expressions available in Ref.~\cite{BIJ02} for circular
orbits. They are computed by implementing the detailed method
described in Ref.~\cite{BI04}. Though algebraically long and involved,
the procedure is fairly algorithmic as explained in~\cite{BIJ02,BI04}.
We thus skip all  those details of computations and list the final results
we need. The 3PN mass
quadrupole $I_{ij}$ is already given in Ref.~\cite{BI04} and its
expression (valid in the frame of the center of mass) is
\begin{equation}
I_{ij}=\nu\,m
\,\Biggl\{\left[A-\frac{24}{7}\,\frac{\nu}{c^5}\,\frac{G^2\,m^2}{r^2}\,\dot{r}\right]\,
x_{< i}x_{j >}+B\,\frac{r^2}{c^2}\,v_{< i}v_{j >} +
2\left[C\,\frac{r\,\dot{r}}{c^2}+\frac{24}{7}
\,\frac{\nu}{c^5}\,\frac{G^2\,m^2}{r}\right]\,x_{< i}v_{j >}
\Biggr\}\,,\label{MQ}
\end{equation}
where the coefficients, up to 3PN order, are
\begin{subequations}\label{coeffsABC}\begin{align}
A &= 1
+\frac{1}{c^2} \left[v^2\, \left( \frac{29}{42}
  - \frac{29\,\nu }{14} \right)+\frac{G\,m}{r}\, \left( -\frac{5}{7}+
  \frac{8 }{7}\,\nu \ \right) \right]\nonumber\\
&\quad+ \frac{1}{c^4}\left[\frac{G\,m}{r}\,v^2\left( \frac{2021}{756}
- \frac{5947 }{756}\,\nu-\frac{4883}{756}\,\nu^2
\right)\right.\nonumber\\&\qquad\quad\quad\left.+\frac{G^2\,m^2}{r^2}\,\left(
- \frac{355}{252} -\frac{953 }{126}\,\nu + \frac{337\,}{252}\,\nu^2
\right)\right.\nonumber\\ &\quad\quad\quad\quad+\left.v^4\,\left(
\frac{253}{504} - \frac{1835 }{504}\,\nu
+\frac{3545}{504}\,\nu^2\right)\right.\nonumber\\&\qquad
\quad\quad\left. +\frac{G\,m}{r}\,\dot{r}^2\left( - \frac{131}{756} +
\frac{907 }{756}\,\nu - \frac{1273}{756}\,\nu^2
\right)\right]\nonumber\\
&\quad+\frac{1}{c^6}\left[v^6\,\left( \frac{4561}{11088} -
\frac{7993}{1584}\,\nu+\frac{117067}{5544}\,\nu^2 -
\frac{328663}{11088}\,\nu^3\right)\right.\nonumber\\
&\left.\qquad\quad\quad+v^4\, \frac{G\,m}{r}\,\left( \frac{307}{77} -
\frac{94475 }{4158}\,\nu+\frac{218411}{8316}\,\nu^2 +
\frac{299857}{8316}\,\nu^3\right)\right.\nonumber\\
&\qquad\quad\quad+\frac{G^3\,m^3}{r^3}\,\left(
\frac{6285233}{207900}+\frac{15502}{385}\,\nu
-\frac{3632}{693}\,\nu^2 + \frac{13289}{8316}\,\nu^3
\right.\nonumber\\
&\left.
\qquad\quad\quad\quad\quad\quad\quad\quad\quad
-\frac{428}{105}\,\ln \left(\frac{r}{r_0}\right) - \frac{44}{3}\,\,\nu
\,\ln \left(\frac{r}{r_0'}\right) \right)\nonumber\\
&\qquad\quad\quad+\frac{G^2\,m^2}{r^2}\,\dot{r}^2\left( -
\frac{8539}{20790}+ \frac{52153 }{4158}\,\nu - \frac{4652}{231}\,\nu^2
-\frac{54121}{5544}\,\nu^3 \right) \,\nonumber\\
&\qquad\quad\quad+\,\frac{G\,m}{r}\,\dot{r}^4\left( \frac{2}{99} -
\frac{1745}{2772}\,\nu +\frac{16319}{5544}\,\nu^2 -
\frac{311\,}{99}\,\nu^3 \right) \,\nonumber\\
&\qquad\quad\quad+\frac{G^2\,m^2}{r^2}\, v^2\left(
\frac{187183}{83160} -\frac{605419 }{16632}\,\nu +
\frac{434909}{16632}\,\nu^2 -\frac{37369}{2772}\,\nu^3
\right)\nonumber\\
&\qquad\quad\quad+\left.\frac{G\,m}{r}\,v^2\,{\dot{r}}^2 \,\left( -
\frac{757}{5544}+\frac{5545 }{8316}\,\nu -
\frac{98311\,}{16632}\,\nu^2 +\frac{153407}{8316}\,\nu^3 \right)
\right]\,,\label{A} \\
B &= \frac{11}{21} - \frac{11}{7}\,\nu\nonumber\\ &
+\frac{1}{c^2}\left[\frac{G m}{r}\,\left( \frac{106}{27} -
\frac{335}{189}\,\nu - \frac{985}{189}\,\nu^2
\right)\right. \nonumber\\ &\quad\qquad\left. +\,v^2\,\left(
\frac{41}{126} - \frac{337 }{126}\,\nu + \frac{733}{126}\,\nu^2
\right)+\dot{r}^2\,\left( \frac{5\,}{63} - \frac{25 }{63}\,\nu +
\frac{25}{63}\,\nu^2 \right)\right] \nonumber\\ &
+\frac{1}{c^4}\,\left[\,v^4\,\left( \frac{1369}{5544} - \frac{19351
}{5544}\,\nu + \frac{45421}{2772}\,\nu^2 - \frac{139999}{5544}\,\nu^3
\right) \right.\nonumber\\
&\qquad\quad +\left.\frac{G^2\, m^2}{r^2}\,\left(-\frac{40716}{1925}-
\frac{10762 }{2079}\,\nu +\frac{62576}{2079}\,\nu^2 -
\frac{24314}{2079}\,\nu^3 \right.\right.\nonumber\\
&\qquad\quad\quad\quad\quad\quad\left.\left.  + \frac{428}{105}\,\ln
\left(\frac{r}{r_0}\right)
\right)\right.\nonumber\\ &\qquad+\,\frac{G m}{r}\,\dot{r}^2\left(
\frac{79}{77}-\frac{5807}{1386}\,\nu + \frac{515}{1386}\,\nu^2
+\frac{8245}{693}\,\nu^3 \right)\nonumber\\ &\qquad+ \frac{G
m}{r}\,v^2\left( \frac{587}{154} - \frac{67933 }{4158}\,\nu+
\frac{25660}{2079}\,\nu^2 +\frac{129781}{4158}\,\nu^3
\right)\nonumber\\ &\qquad+\left. v^2\,\dot{r}^2\,\left(
\frac{115\,}{1386}-\frac{1135}{1386}\,\nu
+\frac{1795}{693}\,\nu^2-\frac{3445}{1386}\,\nu^3
\right)\right]\,,\label{B}\\
C &= -\frac{2}{7}+ \frac{6}{7}\,\nu \nonumber\\ &+
\frac{1}{c^2}\left[v^2\, \left( -\frac{13}{63} + \frac{101 }{63}\,\nu
- \frac{209\,}{63}\,\nu^2 \right) \right.\nonumber\\ &\qquad
\left.+\frac{G\, m}{r}\,\left( -\frac{155}{108} +
\frac{4057}{756}\,\nu + \frac{209}{108}\,\nu^2 \right)\right]
\nonumber\\ &+\frac{1}{c^4}\left[\frac{G\, m}{r}\,v^2 \left( -
\frac{2839}{1386}+ \frac{237893}{16632}\,\nu -
\frac{188063}{8316}\,\nu^2 -\frac{58565}{4158}\,\nu^3
\right)\right.\nonumber\\ &\qquad\qquad+\,\frac{G^2\,
m^2}{r^2}\,\left( -\frac{12587}{41580}+ \frac{406333 }{16632}\,\nu -
\frac{2713}{396}\,\nu^2 +\frac{4441}{2772}\,\nu^3 \right)\nonumber\\
&\qquad\qquad+v^4\,\left( - \frac{457}{2772}+ \frac{6103 }{2772}\,\nu
- \frac{13693}{1386}\,\nu^2+\frac{40687}{2772}\,\nu^3
\right)\nonumber\\ &\qquad\qquad+\left. {\frac{G
m}{r}\,\dot{r}^2\left( \frac{305}{5544} + \frac{3233}{5544}\,\nu -
\frac{8611}{5544}\,\nu^2 - \frac{895}{154}\,\nu^3\right)
}\right]\,.\label{C}
\end{align}\end{subequations}
In the above equation $r_0$ is the length scale appearing in the
definition of the source multipole moments~\cite{BI04} and is the same
as in Eq.~\eqref{Tr}. On the other hand, the different constant $r_0'$
is related to two other length scales $r_1'$ and $r_2'$ (one for each
particle) by $m\ln r_0'=m_1\ln r_1'+m_2 \ln r_2'$, and is specific to
the application of the formalism to point particle systems. It comes
from regularizing the self-field of point particles in the standard
harmonic coordinate system. It is very important to note that the two
length scales $r_1'$ and $r_2'$ are the same as the two scales that
appear in the final expression of the 3PN equations of motion in
standard harmonic coordinates~\cite{BFeom}. The requirement that these
$r_1'$ and $r_2'$ should match with similar scales that appear in 
 the equations of motion
determine, using dimensional regularization, the values of the
Hadamard's regularization constants $\xi$, $\kappa$ and $\zeta$ that
formerly appeared in the 3PN multipole moments~\cite{BIJ02,BI04}. The
regularization constants are thus determined and we
have consistently replaced $\xi$, $\kappa$ and $\zeta$ by their values
known from~\cite{BDEI04,BDEI05}. The constants $r_1'$, $r_2'$ and
hence $r_0'$ are ``unphysical'' in the sense that they can be
arbitrarily changed by a coordinate transformation of the ``bulk''
metric outside the particles~\cite{BFeom}, or, more appropriately
(when considering the renormalisation which follows the dimensional
regularization), by some shifts of the particles' world
lines~\cite{BDE04,BDEI05}.

The 2PN mass octupole and current quadrupole moments for general
orbits are the other non-trivial moments required. They are given by
\begin{subequations}\label{2PNmom}\begin{align}
I_{ijk} &= \nu\,m\, \sqrt{1-4\,\nu }\left\{ \, x_{<ijk>}\left[\,
-1+{{1\over c^2}}\,\left[{{G m}\over{r}}\,\left(\frac{5}{6}-\frac{13\,
\nu }{6}\right)+{v^2}\, \left(-\frac{5}{6}+\frac{19}{6}\,
\nu\right)\right] \right.\right.\nonumber\\ &+ {{1\over{c}^4}}\,
\Bigg[{v^4}\,\Big(-\frac{257}{440}+ \frac{7319}{1320}\, \nu
-\frac{5501}{440}\, {{\nu }^2}\Big)+\frac{G^2\, m^2}{r^2}\,
\Big(\frac{47}{33}+\frac{1591}{132}\, \nu -\frac{235}{66}\,{{\nu
}^2}\Big)\nonumber\\ &+ \left.{{Gm}\over{r}}{{\dot{r}}^2} \,
\Big(\frac{247}{1320}-\frac{531}{440}\,\nu+ \frac{1347}{440}\,{{\nu
}^2}\Big)+\frac{G m}{r}{v^2}\, \Big(-\frac{3853}{1320}+\frac{14257
}{1320}\, \nu +\frac{17371}{1320}\,{{\nu}^2}\Big) \Bigg] \right]
\nonumber\\ &+{x_{<ij}v_{k>}}\,{r\,\dot{r}\over{c}^2}\,\biggl[ 1-2\,
\nu + {1\over{c}^2}\, \Big[{{G m}\over{r}}\,
\Big(\frac{2461}{660}-\frac{8689 }{660}\, \nu-\frac{1389}{220}\, {{\nu
}^2}\Big)\nonumber\\ &+{v^2}\, \Big(\frac{13}{22}-\frac{107}{22}\, \nu
+ \frac{102}{11}\,{{\nu }^2} \Big)\Big] \biggr]+\,{x_{\langle i}v_{jk\rangle }}
{{r^2}\over{c}^2}\,\biggl[-1+2\, \nu + {{1}\over{c}^2}\, \Big[{v^2}\,
\Big(-\frac{61}{110}+\frac{519}{110}\, \nu-\frac{504}{55}\,
{{\nu}^2}\Big)\nonumber\\&+{\dot r^2}\,\Big(\frac{1}{11}-
\frac{4}{11}\,\nu +\frac{3}{11}\, {{\nu }^2}\Big)+{{Gm}\over{r}}\,
\Big(-\frac{1949}{330}-\frac{62 }{165}\, \nu+\frac{483}{55}\,
{{\nu}^2}\Big)\Big] \biggr]\nonumber\\ & \left. +v_{<ijk>} {{{\dot
r}\,{r^3}\over c^4}}\,\Big(-\frac{13}{55}+\frac{52
}{55}\,\nu-\frac{39}{55}\,{{\nu }^2}\Big)\,\right\}\,+{\cal
O}(6)\,,\label{MO}\\ J_{ij}&= m\, \nu\, \sqrt{1-4\,\nu }\,\left\{
\varepsilon_{ab<i} x_{j>a}v_b\left[ -1 + {1\over c^2}\,\left[
{G\,m\over r}\,\left( - \frac{27}{14}- \frac{15}{7}\,\nu \right) +
v^2\,\left( -\frac{13}{28}+ \frac{17}{7}\,\nu \right)\right]
\right.\right.  \nonumber\\ &+{1\over c^4}\,\left[ v^4\,\left( -
\frac{29}{84} + \frac{11}{3}\,\nu - \frac{505}{56}\,{\nu }^2 \right)+
\frac{G^2\,m^2}{ r^2}\left( \frac{43}{252} + \frac{1543 }{126}\,\nu -
\frac{293}{84}\,{\nu }^2 \right) \right.\nonumber\\ &
\left.\left.+\frac{G m}{r} {\dot r^2}\,\left( \frac{5}{252} +
\frac{241}{252}\,\nu + \frac{335}{84}\,{\nu }^2\right) +\frac{G m}{r}
v^2\,\left( - \frac{671}{252} + \frac{1297}{126}\,\nu +
\frac{121}{12}\,{\nu }^2 \right)\right] \right] \nonumber\\ & +
\varepsilon_{ab<i} v_{j>b} x_{a} \, \frac{r\dot r}{c^2}\left[\, -
\frac{5}{28} + \frac{5 }{14}\,\nu +{1\over c^2}\,\left[ v^2\,\left( -
\frac{25}{168}+ \frac{25}{24}\,\nu -\frac{25}{14}\,{\nu }^2
\right)\right.\right. \nonumber\\ &\left.\left.\left.+{G\,m\over r}\,\left(
- \frac{103}{63} - \frac{337}{126}\,\nu + \frac{173}{84}\,{\nu }^2
\right)\right] \right]\right\} \,+\mathcal{O}(6)\,.\label{CQ}
\end{align}\end{subequations}
In the above and what follows, $x_{ijk\cdots}\equiv x_ix_jx_k\cdots$
and $v_{ijk\cdots}\equiv v_iv_jv_k\cdots$, and the brackets $<>$
denote the STF projection. The 1PN moments read as
\begin{subequations}\label{1PNmom}\begin{align}
\label{I4}I_{ijkl} &= \nu\, m \,\left\{x_{<ijkl>}\,\left[1-3\nu+
\frac{1}{c^2} \biggl[ \left(\frac{103}{110}-\frac{147}{22}\nu
+\frac{279}{22}\nu^2\right) v^2\right.\right.\nonumber \\ &-
\left.\left.\left.\left(\frac{10}{11}-\frac{61}{11}\nu
+\frac{105}{11}\nu^2\right) \frac{Gm}{r}\bigg]\right]-
\frac{72}{55}v_{<i}x_{jkl>}\frac{r\,{\dot r}}{c^2}(1-5\nu +5\nu^2)
\right.\right.\nonumber \\ &+
\left.\frac{78}{55}v_{<ij}x_{kl>}\frac{r^2}{c^2}(1-5\nu +5\nu^2)
\right\}+\mathcal{O}(4),\\ J_{ijk} &= \nu m \,\varepsilon_{ab<i}\,\left\{
x_{jk>a}v_b \biggl[ 1-3\nu \right.\nonumber \\ &+
\frac{1}{c^2}\biggl[\left(\frac{41}{90}
-\frac{77}{18}\nu+\frac{185}{18}\nu^2\right) v^2+
\left(\frac{14}{9}-\frac{16}{9}\nu-
\frac{86}{9}\nu^2\right)\frac{Gm}{r}\biggr]\biggr]\nonumber \\
&+\left.\frac{7}{45}v_{jk>b} x_a \frac{r^2}{c^2}(1-5\nu +5\nu^2)\,
+\frac{2}{9}x_{j\underline{a}}v_{k>b}\frac{r\,{\dot r}}{c^2} (1-5\nu
+5\nu^2)\right\}+\mathcal{O}(4)\,.\label{CO}
\end{align}\end{subequations}
[The underlined index $\underline{a}$ means that it should be excluded
from the STF projection.] Finally we  also need
\begin{subequations}\label{Nmom}\begin{align}
I_{ijklm} &= -\nu\, m \,\sqrt{1-4\nu}\,( 1-2\nu)\,x_{<ijklm>}
+\mathcal{O}(2)\,,\label{I5} \\ J_{ijkl} &= -\nu\, m\,\sqrt{1-4\nu}
(1-2\nu) \, \varepsilon_{ab<i}\, x_{jkl>a}v_b +
\mathcal{O}(2)\,,\label{J4}
\end{align}\end{subequations}
as well as $W$, the monopole corresponding to the gauge moments $W_L$,
and which is given by
\begin{equation}\label{W}
W=\frac{1}{3}\,\nu\,m\,r\,\dot{r}+\mathcal{O}(2).
\end{equation}
%
\section{The equations of motion of compact binary systems}\label{EqnOM}
%
\subsection{The equations of motion in standard harmonic coordinates}
\label{eomSH}
The computation of the flux will involve the time derivatives of the
latter source moments. The 3PN accurate flux requires the 3PN
equations of motion for compact binaries which are now
complete~\cite{BFeom,DJSequiv,ABF01,DJSdim,BDE04}. For the present
work, where the multipole moments are computed in standard harmonic
coordinates and reduced to the centre of mass (CM) frame, we require
the 3PN accurate equation of motion (or acceleration) in the CM frame
associated with the standard harmonic gauge. This was computed
in~\cite{BI03CM} and given as
\begin{equation}\label{EOM}
a^i=\frac{d v^i}{dt}=-\frac{G m}{r^2}\Big[(1+P)\,n^i + Q\,v^i \Big]+
\mathcal{O}(7)\;,
\end{equation}
where the coefficients $P$ and $Q$ are:
\begin{subequations}\label{EOMcoeff}\begin{align}
P &= \frac{1}{c^2}\left\{-\frac{3\,\dot{r}^2\,\nu}{2} + v^2 +
3\,\nu\,v^2-\frac{G m}{r}\left(4 +2\,\nu \right)\right\}\nonumber\\ &+
\frac{1}{c^4}\left\{\frac{15\,\dot{r}^4\,\nu}{8} -
\frac{45\,\dot{r}^4\,\nu^2}{8} - \frac{9\,\dot{r}^2\,\nu\,v^2}{2} +
6\,\dot{r}^2\,\nu^2\,v^2 + 3\,\nu\,v^4 - 4\,\nu^2\,v^4\right.
\nonumber\\ &\qquad + \left.\frac{G m}{r}\left( -2\,\dot{r}^2 -
25\,\dot{r}^2\,\nu - 2\,\dot{r}^2\,\nu^2 - \frac{13\,\nu\,v^2}{2} +
2\,\nu^2\,v^2 \right)\right. \nonumber\\ &\qquad
+\left.\frac{G^2 m^2}{r^2}\,\left( 9 + \frac{87\,\nu}{4}
\right)\right\}\nonumber\\&+\frac{1}{c^5}\left\{-
\frac{24\,\dot{r}\,\nu\,v^2}{5}\frac{G m}{r}-\frac{136\,\dot{r}\,\nu}{15}
\frac{G^2 m^2}{r^2}\right\}\nonumber\\ &+
\frac{1}{c^6}\left\{-\frac{35\,\dot{r}^6\,\nu}{16} +
\frac{175\,\dot{r}^6\,\nu^2}{16} -
\frac{175\,\dot{r}^6\,\nu^3}{16}+\frac{15\,\dot{r}^4\,\nu\,v^2}{2}
\right.\nonumber\\&\qquad -
\left. \frac{135\,\dot{r}^4\,\nu^2\,v^2}{4} +
\frac{255\,\dot{r}^4\,\nu^3\,v^2}{8} -
\frac{15\,\dot{r}^2\,\nu\,v^4}{2} +
\frac{237\,\dot{r}^2\,\nu^2\,v^4}{8} \right.\nonumber\\ &\qquad
-\left. \frac{45\,\dot{r}^2\,\nu^3\,v^4}{2} + \frac{11\,\nu\,v^6}{4} -
\frac{49\,\nu^2\,v^6}{4} + 13\,\nu^3\,v^6 \right.\nonumber\\ &\qquad +
\left.\frac{G m}{r}\left( 79\,\dot{r}^4\,\nu -
\frac{69\,\dot{r}^4\,\nu^2}{2} - 30\,\dot{r}^4\,\nu^3 -
121\,\dot{r}^2\,\nu\,v^2 + 16\,\dot{r}^2\,\nu^2\,v^2
\right.\right.\nonumber\\&\qquad\qquad\quad~ +\left.\left.
20\,\dot{r}^2\,\nu^3\,v^2+\frac{75\,\nu\,v^4}{4} + 8\,\nu^2\,v^4 -
10\,\nu^3\,v^4 \right)\right.\nonumber\\ &\qquad + \left.
\frac{G^2 m^2}{r^2}\,\left( \dot{r}^2 + \frac{32573\,\dot{r}^2\,\nu}{168}
+ \frac{11\,\dot{r}^2\,\nu^2}{8} - 7\,\dot{r}^2\,\nu^3 +
\frac{615\,\dot{r}^2\,\nu\,\pi^2}{64} - \frac{26987\,\nu\,v^2}{840}
\right.\right.\nonumber\\&\qquad\qquad\quad~ +\left.\left. \nu^3\,v^2
- \frac{123\,\nu\,\pi^2\,v^2}{64} - 110\,\dot{r}^2\,\nu\,\ln
\Big(\frac{r}{r'_0}\Big) + 22\,\nu\,v^2\,\ln \Big(\frac{r}{r'_0}\Big)
\right)\right.\nonumber\\&\qquad +\left.\frac{G^3 m^3}{r^3}\left( -16 -
\frac{437\,\nu}{4} - \frac{71\,\nu^2}{2} + \frac{41\,\nu\,{\pi
}^2}{16} \right)\right\}\;,\label{EOMa}\\ Q
&=\frac{1}{c^2}\Big\{-4\,\dot{r} + 2\,\dot{r}\,\nu\Big\}
\nonumber\\&+\frac{1}{c^4}\left\{\frac{9\,\dot{r}^3\,\nu}{2} +
3\,\dot{r}^3\,\nu^2 -\frac{15\,\dot{r}\,\nu\,v^2}{2} -
2\,\dot{r}\,\nu^2\,v^2\right.\nonumber\\ &\qquad +
\left.\frac{G m}{r}\left( 2\,\dot{r} + \frac{41\,\dot{r}\,\nu}{2} +
4\,\dot{r}\,\nu^2 \right)\right\}\nonumber\\&+\frac{1}{c^5}\left\{
\frac{8\,\nu\,v^2}{5}\frac{G m}{r}+\frac{24\,\nu}{5}
\frac{G^2 m^2}{r^2}\right\}\nonumber\\ &+
\frac{1}{c^6}\left\{-\frac{45\,\dot{r}^5\,\nu}{8} +
15\,\dot{r}^5\,\nu^2 + \frac{15\,\dot{r}^5\,\nu^3}{4} +
12\,\dot{r}^3\,\nu\,v^2 \right.\nonumber\\&\qquad
-\left. \frac{111\,\dot{r}^3\,\nu^2\,v^2}{4}
-12\,\dot{r}^3\,\nu^3\,v^2 -\frac{65\,\dot{r}\,\nu\,v^4}{8} +
19\,\dot{r}\,\nu^2\,v^4 + 6\,\dot{r}\,\nu^3\,v^4
\right.\nonumber\\&\qquad\left. + \frac{G m}{r}\left(
\frac{329\,\dot{r}^3\,\nu}{6} + \frac{59\,\dot{r}^3\,\nu^2}{2} +
18\,\dot{r}^3\,\nu^3 - 15\,\dot{r}\,\nu\,v^2 - 27\,\dot{r}\,\nu^2\,v^2
- 10\,\dot{r}\,\nu^3\,v^2 \right) \right.\nonumber\\&\qquad
+\left.\frac{G^2 m^2}{r^2}\,\left( -4\,\dot{r} -
\frac{18169\,\dot{r}\,\nu}{840} + 25\,\dot{r}\,\nu^2 +
8\,\dot{r}\,\nu^3 - \frac{123\,\dot{r}\,\nu\,\pi^2}{32}
\right.\right.\nonumber\\&\qquad\qquad\quad~\left.\left. +
44\,\dot{r}\,\nu\,\ln \Big(\frac{r}{r'_0}\Big)
\right)\right\}\;.\label{EOMb}
\end{align}\end{subequations}
Recall that there was initially a regularization ambiguity constant
denoted $\lambda$ in~\cite{BFeom}, which has been replaced here by its
uniquely determined value
$\lambda=-\frac{1987}{3080}$~\cite{BDE04}. On the other hand the
constant $r'_0$ is the \textit{same} as the one in the 3PN quadrupole
moment~\eqref{MQ}--\eqref{coeffsABC}.
\subsection{The modified harmonic coordinates (without logarithms)}\label{eomMH}
The standard harmonic (hereafter SH) coordinate system used till now
is useful for analytical algebraic checks, but contains
gauge-dependent logarithmic terms that are not very convenient in
numerical calculations. More importantly, in the presence of the
logarithmic terms the simple generalized quasi-Keplerian
representation (reviewed in Sect.~\ref{QKrepr}) is not possible
impeding the process of averaging the flux over the orbital
period. Consequently, it is useful to have the expression for the
energy flux in a modified harmonic (MH) coordinate system without
logarithms like the one explicitly used in~\cite{MW03} (we shall
alternatively use ADM type coordinates which are also free of such
logarithms at 3PN). This will require us to re-express the
instantaneous expressions for the energy flux [given by
Eqs.~\eqref{3PNEF-inst} below] in terms of corresponding variables in
the MH or ADM coordinate systems. We provide in this section the
definition of the MH coordinate system.

Consider the coordinate transformation
$x'^\mu=x^\mu+\varepsilon^\mu(x)$ which removes the logarithms $\ln
(r/r'_0)$ at the level of the equations of motion as discussed in
Ref.~\cite{BFeom}. It is given by
\begin{equation}\label{epsmu}
\varepsilon_\mu = \frac{22}{3}
\frac{G^2\,m_1\,m_2}{c^6}\,\partial_\mu\,\left[
\frac{G\,m_2}{r_1}\ln\left(\frac{r}{r'_2}\right)+\frac{G\,m_1}{r_2}
\ln\left(\frac{r}{r'_1}\right)\right]\,,
\end{equation}
where $r_1=\vert\mathbf{x}-\mathbf{y}_1\vert$ and
$r_2=\vert\mathbf{x}-\mathbf{y}_2\vert$ are the distances to the two
particles with trajectories $y_1^i(t)$ and $y_2^i(t)$, and where
$r=\vert\mathbf{y}_1-\mathbf{y}_2\vert$ is their relative
distance. Following~\cite{BDE04} the logarithms can be equivalently
removed by the shifts (sometimes also called the ``contact''
transformations) of the particle world-lines induced by the change of
coordinates, namely
\begin{subequations}\label{shifts}
\begin{align}
{y'}_1^i &= y_1^i + \xi_1^i \,,\\ {y'}_2^i &= y_2^i + \xi_2^i \,.
\end{align}
\end{subequations}
The equalities here are \textit{functional} relations, \textit{i.e.}
the two sides of the equations are evaluated at the \textit{same}
coordinate time say, $t$. The spatial shifts $\xi_1^i$ and $\xi_2^i$ of
the two world-lines are related to the coordinate transformation
restricted to the world-lines [denoted 
$\varepsilon_1^\mu(t)\equiv  
\varepsilon^\mu(t,y_1(t)) $ 
and
$\varepsilon_2^\mu(t)\equiv  
\varepsilon^\mu(t,y_2(t))] $ by 
\begin{subequations}\label{shifts1}
\begin{align}
\xi_1^i=\varepsilon_1^i-\frac{v_1^i}{c}\,\varepsilon_1^0+
\mathcal{O}\left(\varepsilon^2\right)\,,\\
\xi_2^i=\varepsilon_2^i-\frac{v_2^i}{c}\,\varepsilon_2^0+
\mathcal{O}\left(\varepsilon^2\right)\,,
\end{align}
\end{subequations}
where $v_1^i=dy_1^i/dt$ and $v_2^i=dy_2^i/dt$ are the coordinate
velocities.
The latter relations are valid at linear order in
$\varepsilon_1^\mu $ and 
$\varepsilon_2^\mu $. 
Now the coordinate transformation~\eqref{epsmu} is at 3PN
order so we have $\varepsilon^0=\mathcal{O}(7)$ and
$\varepsilon^i=\mathcal{O}(6)$. Hence, we see from \eqref{shifts1} that
 at the 3PN order the shifts
simply agree with the spatial components of the coordinate
transformation,
\begin{subequations}\label{shifts2}
\begin{align}
\xi_1^i=\varepsilon_1^i + \mathcal{O}\left(8\right)\,,\\
\xi_2^i=\varepsilon_2^i + \mathcal{O}\left(8\right)\,.
\end{align}
\end{subequations}
They are readily obtained from Eq.~\eqref{epsmu} as
\begin{subequations}\label{shifts3}
\begin{align}
\xi_1^i &=- \frac{22}{3}
\frac{G^3\,m_1^2\,m_2}{c^6\,r^2}\,n^i\,\ln\left(\frac{r}{r_1'}\right)
+\mathcal{O}(8)\,,\\ \xi_2^i &= \frac{22}{3}
\frac{G^3\,m_1\,m_2^2}{c^6\,r^2}\,n^i\,\ln\left(\frac{r}{r_2'}\right)
+\mathcal{O}(8)\,.
\end{align}
\end{subequations}

Under the shifts of world-lines the accelerations of the particles are
changed by the amounts $\delta_\xi a_1^i$ and $\delta_\xi a_2^i$
(\textit{i.e.} such that the functional equalities
${a'}_{1,2}^i=a_{1,2}^i+\delta_\xi a_{1,2}^i$ hold) given by
\begin{subequations}\label{deltaa12}
\begin{align}
\delta_{\xi}a^i_1 = \frac{d^2\xi_1^i}{dt^2} -
\left(\xi_1^j-\xi_2^j\right)\partial_{ij}\left(\frac{G\,m_2}{r}\right)
+ \mathcal{O}(8)\,,\\ \delta_{\xi}a^i_2 = \frac{d^2\xi_2^i}{dt^2} +
\left(\xi_1^j-\xi_2^j\right)\partial_{ij}\left(\frac{G\,m_1}{r}\right)
+ \mathcal{O}(8)\,.
\end{align}
\end{subequations}
where the second terms come from re-expressing the gravitational force
-- gradient of the Newtonian potential -- in terms of the new
trajectories~\eqref{shifts}. The relative acceleration $a^i \equiv a_1^i -
a_2^i$ is changed by the amount
\begin{equation}\label{delta-accn}
\delta_{\xi}a^i = \frac{d^2\xi_{12}^i}{dt^2} -
\xi_{12}^j\,\partial_{ij}\left(\frac{G\,m}{r}\right) +
\mathcal{O}(8)\,,
\end{equation}
where $m\equiv m_1+m_2$ and $\xi_{12}^i\equiv\xi_1^i-\xi_2^i$.\footnote{This means that $x^i_{\rm MH}=x^i_{\rm SH} +\xi^i_{12}$.} An easy
calculation shows that the change in the relative acceleration
associated with the shifts~\eqref{shifts3} is
\begin{align}\label{accn-BItoMW}
\delta_\xi a^i&=\frac{G^3\,m^3\,\nu}{c^6\,r^4}\left\{
\biggl[\left(-110 \dot{r}^2 +22 v^2\right)n^i+44 \dot{r} \,v^i\biggr]
\ln\left(\frac{r}{r_0'}\right) \right.\nonumber\\
&\left.\qquad\qquad+\left(\frac{176}{3}\dot{r}^2
-\frac{22}{3}v^2+\frac{22}{3}\frac{G\,m}{r}\right)n^i
-\frac{44}{3}\dot{r}\,v^i\right\} + \mathcal{O}(8)\,.
\end{align}
Adding the above shift to the expression for the relative acceleration
in SH coordinates as given by Eqs.~\eqref{EOM}--\eqref{EOMcoeff},
yields the expression for the acceleration in MH coordinates. Since
$a'^i=a^i+\delta_\xi a^i$ is a functional identity, the resulting MH
acceleration is obtained as a function of the ``dummy'' variables
denoted $v^2$, $\dot{r}$ and $r$. Evidently these variables are to be
interpreted as the natural variables describing the binary motion in
MH coordinates.\footnote{To avoid making the notation too heavy we do
not add a subscript MH on the variables $v^2$, $\dot{r}$ and $r$. 
In the following our
notation may not always be completely consistent but should be clear from
the context.} As expected, the logarithms in Eq.~\eqref{accn-BItoMW}
exactly cancel the logarithms in the SH
acceleration~\eqref{EOM}--\eqref{EOMcoeff}. Some 3PN coefficients in
the EOM are also modified and the final result agrees with that
displayed in Ref.~\cite{MW03} (see also~\cite{Bliving}).

For completeness we note also that the above shifts will modify the
3PN conserved energy of the binary (associated with the conservative
part of the 3PN equations of motion) by the amount
\begin{equation}\label{deltaE}
\delta_{\xi}E =- m_1\,v_1^i\,\frac{d\xi_1^i}{dt} -
m_2\,v_2^i\,\frac{d\xi_2^i}{dt} +
\xi_{12}^i\,\partial_i\left(\frac{G\,m_1\,m_2}{r}\right) +
\mathcal{O}(8)\,.
\end{equation}
For the case at hand with the shifts~\eqref{shifts3} and in the
center-of-mass frame we find
\begin{equation}\label{deltaEres}
\delta_{\xi}E = \frac{22}{3}\frac{G^3\,m^4\,\nu^2}{c^6\,r^3}\left\{
\biggl[\frac{G\,m}{r}-3\dot{r}^2 +v^2\biggr]\ln\left(\frac{r}{r_0'}\right)
+\dot{r}^2\right\} + \mathcal{O}(8)\,.
\end{equation}
Comparing with the 3PN energy in SH coordinates as given by Eq.~(4.8)
in~\cite{BI03CM}, we see that the logarithms $\ln(r/r_0')$ are also
cancelled in the expression for the energy by going to the MH coordinates.
%
\section{The instantaneous part of the 3PN energy flux}
\label{EF-Sharm} 
%
Using the multipole moments given in Eqs.~\eqref{MQ}--\eqref{Nmom},
one computes the required time derivatives with the help of the
equations of motion~\eqref{EOM} and obtains the instantaneous part of
the energy flux as defined by Eq.~\eqref{inst}. Here we are working in
SH coordinates, in which the equations of motion are given by
Eqs.~\eqref{EOM}--\eqref{EOMcoeff}. In the next section we consider the case of
alternative coordinate systems. The hereditary
part computed in Paper~I will be added after the process of averaging
over one orbit (this contribution is the same in all alternative
coordinate systems considered in this paper).
Though lengthy the computation of the different parts
constituting the instantaneous terms in the energy flux at 3PN order
is straightforward.\footnote{In order to perform some independent
checks on the long and involved algebra, we have found it expeditious
to make two computations using the two harmonic coordinate systems: SH
containing the (gauge dependent) log terms \textit{\`a
la}~\cite{BI03CM} and MH without log terms as in
Refs.~\cite{MW03,Bliving}.} We write the result as
\begin{equation}
\label{EF} \mathcal{F}_\mathrm{inst} = \mathcal{F}^\mathrm{N}_\mathrm{inst}+
\mathcal{F}^\mathrm{1PN}_\mathrm{inst}+ \mathcal{F}^{\rm
2PN}_\mathrm{inst}+\mathcal{F}^\mathrm{2.5PN}_\mathrm{inst}+
\mathcal{F}^\mathrm{3PN}_\mathrm{inst} +\mathcal{O}(7)\,,
\end{equation}
and find that the various PN pieces are given by
\begin{subequations} \label{3PNEF-inst}
\begin{align}
\mathcal{F}^\mathrm{N}_\mathrm{inst}&={\frac{32}{5} \frac{G^3\,
m^4\,\nu^2}{c^5\, r^4} \left\{v^2-\frac{11}{12} \dot{r}^2\right\}}\,,
\label{0PNEF}\\
\mathcal{F}^\mathrm{1PN}_\mathrm{inst}&={\frac{32}{5}\frac{G^3\,
m^4\,\nu^2}{c^7\, r^4}\left\{v^4 \left(\frac{785}{336}-\frac{71}{28} \nu
\right)+{\dot r}^2\,v^2 \left(-\frac{1487}{168}+\frac{58}{7} \nu
\right)\right.}\nonumber\\& {+\frac{G\,m}{r} v^2
\left(-\frac{170}{21}+\frac{10}{21} \nu \right)+{\dot r}^4
\left(\frac{687}{112}-\frac{155}{28} \nu \right)}\nonumber\\&
{\left.+\frac{G\,m}{r} {\dot r}^2 \left(\frac{367}{42}-\frac{5}{14} \nu
\right)+\frac{G^2\,m^2}{r^2} \left(\frac{1}{21}-\frac{4}{21} \nu
\right)\right\}}\,,
\label{1PNEF}\\
\mathcal{F}^\mathrm{2PN}_\mathrm{inst}&={\frac{32}{5}\frac{G^3\,
m^4\,\nu^2}{c^9\, r^4}\left\{v^6 \left(\frac{47}{14}-\frac{5497}{504}
\nu +\frac{2215}{252} \nu ^2\right)\right.}\nonumber\\& {+\dot{r}^2 v^4
\left(-\frac{573}{56}+\frac{1713}{28} \nu -\frac{1573}{42} \nu
^2\right)}\nonumber\\& {+\frac{G\, m}{r} v^4
\left(-\frac{247}{14}+\frac{5237}{252} \nu -\frac{199}{36} \nu
^2\right)}\nonumber\\& {+\dot{r}^4 v^2
\left(\frac{1009}{84}-\frac{5069}{56} \nu +\frac{631}{14} \nu
^2\right)}\nonumber\\& {+\frac{G\, m}{r} \dot{r}^2 v^2
\left(\frac{4987}{84}-\frac{8513}{84} \nu +\frac{2165}{84} \nu
^2\right)}\nonumber\\& {+\frac{G^2\, m^2}{r^2} v^2
\left(\frac{281473}{9072}+\frac{2273}{252} \nu +\frac{13}{27} \nu
^2\right)}\nonumber\\& {+\dot{r}^6
\left(-\frac{2501}{504}+\frac{10117}{252} \nu -\frac{2101}{126} \nu
^2\right)}\nonumber\\& {+\frac{G\, m}{r} \dot{r}^4
\left(-\frac{5585}{126}+\frac{60971}{756} \nu -\frac{7145}{378} \nu
^2\right)}\nonumber\\& {+\frac{G^2\, m^2}{r^2} \dot{r}^2
\left(-\frac{106319}{3024}-\frac{1633}{504} \nu -\frac{16}{9} \nu
^2\right)}\nonumber\\& {\left.+\frac{G^3\, m^3}{r^3}
\left(-\frac{253}{378}+\frac{19}{7} \nu -\frac{4}{27} \nu
^2\right)\right\}}\,,
\label{2PNEF}\\
\mathcal{F}^\mathrm{2.5PN}_\mathrm{inst}&=
\frac{32}{5}\,\frac{G^3\,m^4\,\nu^2\,}{c^{10}\,r^4}
\left\{\dot{r}\,\nu\left( -\frac{12349}{210}\, \frac{G\,m}{r} v^4
+\frac{4524}{35} \frac{G\,m}{r} v^2\dot{r}^2 -\frac{2753}{126}
\frac{G^2\, m^2}{r^2} v^2 \right.\right.\nonumber\\
&\left.\left.-\frac{985}{14}\, \frac{G\,m}{r} \dot{r}^4 +
\frac{13981}{630}\, \frac{G^2\, m^2}{r^2} \dot {r}^2 -\frac{1}{315}
\frac{G^3\, m^3}{r^3} \right)\right\}\,,
\label{2p5PNEF}\\
\mathcal{F}^\mathrm{3PN}_\mathrm{inst}&={\frac{32}{5}\frac{G^3\,
m^4\,\nu^2}{c^{11}\, r^4}\left\{v^8
\left(\frac{80315}{14784}-\frac{694427}{22176} \nu
+\frac{604085}{11088} \nu ^2-\frac{16985}{462} \nu
^3\right)\right.}\nonumber\\& {+\dot{r}^2 v^6
\left(-\frac{31499}{1008}+\frac{1119913}{5544} \nu -\frac{44701}{132}
\nu ^2+\frac{38725}{231} \nu ^3\right)}\nonumber\\& {+\frac{G\, m}{r}
v^6 \left(-\frac{61669}{3696}+\frac{95321}{1008} \nu
-\frac{955013}{11088} \nu ^2+\frac{47255}{1386} \nu
^3\right)}\nonumber\\& {+\dot{r}^4 v^4
\left(\frac{204349}{2464}-\frac{3522149}{7392} \nu
+\frac{2354753}{3696} \nu ^2-\frac{109447}{462} \nu
^3\right)}\nonumber\\& {+\frac{G\, m}{r} \dot{r}^2 v^4
\left(\frac{136695}{1232}-\frac{202693}{336} \nu +\frac{744377}{1232}
\nu ^2-\frac{931099}{5544} \nu ^3\right)}\nonumber\\& {+\frac{G^2\,
m^2}{r^2} v^4 \left(\frac{598614941}{2494800}-\frac{856}{35}
\ln\left(\frac{r}{r_0}\right)\right.}\nonumber\\&
{\left.+\left[\frac{39896}{2079}-\frac{369}{64} \pi ^2\right] \nu
+\frac{1300907}{33264} \nu ^2-\frac{161783}{24948} \nu
^3\right)}\nonumber\\& {+\dot{r}^6 v^2
\left(-\frac{1005979}{11088}+\frac{2589599}{5544} \nu
-\frac{1322141}{2772} \nu ^2+\frac{90455}{693} \nu
^3\right)}\nonumber\\& {+\frac{G\, m}{r} \dot{r}^4 v^2
\left(-\frac{715157}{3696}+\frac{35158037}{33264} \nu
-\frac{3672143}{3696} \nu ^2+\frac{871025}{4158} \nu
^3\right)}\nonumber\\& {+\frac{G^2 m^2}{r^2} \dot{r}^2 v^2
\left(-\frac{35629009}{37800}+\frac{3424}{35}
\ln\left(\frac{r}{r_0}\right)\right.}\nonumber\\&
{\left.+\left[-\frac{150739}{1232}+\frac{861}{32} \pi ^2\right] \nu
-\frac{453247}{1848} \nu ^2+\frac{496081}{8316} \nu
^3\right)}\nonumber\\& {+\frac{G^3\, m^3}{r^3} v^2
\left(-\frac{24608492}{155925}+\frac{856}{105}
\ln\left(\frac{r}{r_0}\right)\right.}\nonumber\\&
{\left.+\left[-\frac{6356291}{22680}+\frac{44}{3}
\ln\left(\frac{r}{r'_0}\right)+\frac{451}{64} \pi ^2\right] \nu
+\frac{3725}{462} \nu ^2-\frac{841}{2268} \nu ^3\right)}\nonumber\\&
{+\dot{r}^8 \left(\frac{1507925}{44352}-\frac{20365}{126} \nu
+\frac{687305}{5544} \nu ^2-\frac{32755}{1386} \nu
^3\right)}\nonumber\\& {+\frac{G\, m}{r} \dot{r}^6
\left(\frac{5476951}{55440}-\frac{671765}{1232} \nu
+\frac{5205019}{11088} \nu ^2-\frac{860477}{11088} \nu
^3\right)}\nonumber\\& {+\frac{G^2\, m^2}{r^2} \dot{r}^4
\left(\frac{115627817}{166320}-\frac{214}{3}
\ln\left(\frac{r}{r_0}\right)\right.}\nonumber\\&
{\left.+\left[\frac{42671}{792}-\frac{697}{32} \pi ^2\right] \nu
+\frac{1099355}{4752} \nu ^2-\frac{825331}{16632} \nu
^3\right)}\nonumber\\& {+\frac{G^3\, m^3}{r^3} \dot{r}^2
\left(\frac{3202601}{23100}-\frac{1712}{315}
\ln\left(\frac{r}{r_0}\right)\right.}\nonumber\\&
{\left.+\left[\frac{6220199}{22680}-\frac{88}{9}
\ln\left(\frac{r}{r'_0}\right)-\frac{1763}{192} \pi ^2\right] \nu
+\frac{57577}{1848} \nu ^2-\frac{43018}{6237} \nu
^3\right)}\nonumber\\& {\left.+\frac{G^4\, m^4}{r^4}
\left(\frac{37571}{8316}-\frac{14962}{891} \nu -\frac{3019}{594} \nu
^2-\frac{866}{6237} \nu ^3\right)\right\}}\,. \label{3PNEF}
\end{align}
\end{subequations}
The new results are the instantaneous terms at 2.5PN and 3PN
orders. Up to 2PN order, all the terms match with those obtained in
Refs.~\cite{PM63,BS89,GI97}. As one may notice, the 2.5PN terms in the
above equation are all proportional to ${\dot r}$ and hence are zero
for the circular orbit case in agreement with the result
of~\cite{B96}. The $\dot{r}$ dependence of the 2.5PN terms is
important when we discuss their orbital average in
Sect.~\ref{avgE}. The 3PN terms provide the generalization of the
circular orbit results in Ref.~\cite{BIJ02}. As expected, the constant
$r_0$ present in the expression of the mass quadrupole moment appears
in the final expression for the 3PN flux (the presence of $r_0'$ is of
a different type and is dealt with in the next Section). The
dependence of the instantaneous terms on the scale $r_0$ should
exactly cancel a similar contribution coming from the tail terms as
determined in Paper~I. This cancellation has already been checked for
circular orbits in~\cite{BIJ02} and we shall prove this cancellation
for quasi-elliptical orbits in Sect.~\ref{avgE}.
%
\section{The 3PN energy flux in alternative coordinates}
\label{EF-alt} 
%
The dependence on $r_0'$ of the result~\eqref{EF}--\eqref{3PNEF-inst}
is due to our use of the SH coordinate system. For circular orbits, it
was shown~\cite{BIJ02} that this $r_0'$ dependence disappears when the
total flux is expressed in terms of the gauge invariant parameter
$x=(G m\omega/c^3)^{2/3}$ related to the GW frequency. In the general orbit case we shall transform away the dependence on $r_0'$ by going to
different coordinate systems such as the MH coordinates studied in
Sect.~\ref{eomMH}. Subsequently we shall average the energy flux over an orbital
period and exhibit alternative representations
of the energy flux for elliptical orbits. In particular some of these
are in terms of gauge invariant variables
related to those suggested in Ref.~\cite{MGS04}.
\subsection{The modified harmonic coordinates}\label{MH}
We now provide the energy flux $\mathcal{F}$ in the modified harmonic
(MH) system avoiding the appearance of the logarithms $\ln(r/r'_0)$ at
3PN order and which has been introduced in Sect.~\ref{eomMH}. First we
notice that $\mathcal{F}$ is a function of the ``natural'' variables
$r$, $\dot{r}$ and $v^2$ [see Eqs.~\eqref{3PNEF-inst}], and is a
\textit{scalar}, therefore it satisfies, under the shifts of these
variables defined by~\eqref{shifts},
\begin{equation}\label{IijMH}
\mathcal{F}[r,\dot{r},v^2]=\mathcal{F}'[r',\dot{r}',v'^2]\,.
\end{equation}
This means that we shall have the \textit{functional} equality
$\mathcal{F}'=\mathcal{F} +\delta_\xi\mathcal{F}$ in which
\begin{equation}\label{deltaF}
\delta_\xi\mathcal{F}=- \delta_\xi r\frac{\partial \mathcal{F}}{\partial
r}-\delta_\xi \dot{r}\frac{\partial \mathcal{F}}{\partial \dot{r}}-\delta_\xi
v^2\frac{\partial \mathcal{F}}{\partial v^2} + \mathcal{O}\left(\xi^2\right)\,,
\end{equation}
where 
\begin{equation}\left.\begin{array}{rll}
\delta_\xi r &=& n^i\,\xi_{12}^i\\[0.2cm] \delta_\xi \dot{r} &=&
\displaystyle n^i\,\frac{d \xi_{12}^i}{d t} + \frac{v^i-\dot{r}
\,n^i}{r}\,\xi_{12}^i\\[0.2cm] \delta_\xi v^2 &=& \displaystyle 2
v^i\,\frac{d \xi_{12}^i}{d t}\end{array}\right\} +
\mathcal{O}(\xi^2)\,.\label{transf}
\end{equation}
(Recall $\xi_{12}^i\equiv\xi_{1}^i-\xi_{2}^i$.) Since the previous
formulas are at linear order in the shifts and we are interested in
the 3PN approximation, they are valid for any shifts at the 3PN order
(case of the MH coordinates) and also at the 2PN order like the ones
associated with the passage to ADM coordinates -- in the latter
case, the error will be at 4PN order.

In the case of the MH shifts, which start at the 3PN order, one can
make an alternative computation of the modification of the energy
flux. Indeed the only modification \textit{vis \`a vis} the
calculation in standard harmonic (SH) coordinates is the one related
to the mass quadrupole moment which must be computed to 3PN
accuracy. Under the shifts of the particles' trajectories $y_{1,2}^i$
as given by~\eqref{shifts}, the mass quadrupole moment $I_{ij}$, which
equals $I_{ij}=m_1 \,y_1^{<ij>}+m_2 \,y_2^{<ij>}+\mathcal{O}(2)$ in
the Newtonian approximation, is shifted by the amount
($1\leftrightarrow 2$ meaning the same term but for the other
particle)
\begin{equation}\label{deltaIij}
\delta_\xi I_{ij} = 2 m_1\,y_1^{<i}\xi_1^{j>} + 1\leftrightarrow 2 +
\mathcal{O}(8)\,,
\end{equation}
where the remainder $\mathcal{O}(8)$ comes from the 1PN corrections in
the quadrupole moment coupled with the 3PN shifts. Using the explicit
expressions of these shifts in~\eqref{shifts3} we find, in the
center-of-mass frame,
\begin{equation}\label{deltaIijexpl}
\delta_\xi I_{ij} =- \frac{44}{3}
\frac{G^3 m^4\,\nu^2}{r^3} \ln\left(\frac{r}{r_0'}\right)x^{<ij>} + \mathcal{O}(8)\,,
\end{equation}
where $r_0'$ is given by $m\ln r_0'=m_1\ln r_1'+m_2\ln r_2'$. This
modification of the quadrupole moment is seen to exactly cancel the
$\ln(r/r_0')$ dependence of the mass quadrupole moment in SH
coordinates as given by~\eqref{MQ}--\eqref{coeffsABC}. Thus in the MH
gauge the $r_0'$ dependence of the mass quadrupole moment vanishes as
expected. The rest of the expression of the moment remains exactly the
same as in SH coordinates, Eqs.~\eqref{MQ}--\eqref{coeffsABC}, and will
not be repeated here.

Next we must take into account the fact that when computing the third
time derivative of the quadrupole moment, needed in the expression of
the flux, the acceleration in MH coordinates is modified. 
We get
\begin{equation}
\delta_\xi\left(\dddot{I}_{ij}\right) = \frac{d^3}{d
t^3}\left(\delta_\xi I_{ij}\right) - 2 m_1\left[3 v_1^{<i}\,\delta_\xi
a_1^{j>} + y_1^{<i}\,\frac{d\delta_\xi a_1^{j>}}{d t}\right] +
1\leftrightarrow 2 + \mathcal{O}(8)\,,
\end{equation}
where the dots mean the time derivative. The first term is the third
time derivative of the direct modification of the quadrupole moment,
Eqs.~\eqref{deltaIij}--\eqref{deltaIijexpl}, and the extra terms come
from the modification of the accelerations which are given
by~\eqref{deltaa12}. On the other hand, all the other contributions coming
from the higher multipole moments and their derivatives remain
unchanged. We then find
\begin{equation}
\delta_\xi\mathcal{F} =-
\frac{2G}{5c^5}\,\left[\dddot{I}_{ij}\,\delta_\xi\left(\dddot{I}_{ij}\right)
+ \mathcal{O}(8)\right]\,.
\end{equation}
With the explicit expression of the shifts one finally obtains the
modification of the 3PN energy flux in the MH coordinates as (thus, ${\cal F}_{\rm MH}={\cal F}_{\rm SH} +\delta_\xi{\cal F}$)
\begin{equation} \label{EF-ShtoMh}
\delta_\xi\mathcal{F} =- \frac{1408}{15}\,\frac{G^6 \,m^7
\,\nu^3}{c^{11}\,r^7}\left[ \left(v^2 - \frac{2}{3} \dot{r}^2
\right)\ln\left(\frac{r}{r'_0}\right) - \frac{\dot{r}^2}{12} +
\mathcal{O}(2)\right]\,.
\end{equation}
[Of course the result agrees with the one we would obtain from using
directly Eqs.~\eqref{deltaF}--\eqref{transf}.]
\subsection{The ADM coordinates}\label{EF-Mharm}
Many related numerical relativity studies are in ADM (or ADM-type)
coordinates and hence for future applications we wish to provide the
explicit expression for the 3PN energy flux in ADM coordinates. To
transform the energy flux we require the shift or contact
transformation of the trajectories connecting the SH coordinates (with
log terms) and the ADM coordinates. We recall that the ADM and SH
coordinate systems agree at 1PN order inclusively, so that the contact
transformation is composed of 2PN and 3PN terms. Hence the calculation
is more involved than for the MH coordinates (for which only the
modification of the quadrupole moment $I_{ij}$ played a role), and we
must come back to the general
formulas~\eqref{deltaF}--\eqref{transf}. Note that the remainder
$\mathcal{O}\left(\xi^2\right)$ in Eqs.~\eqref{deltaF}--\eqref{transf}
is of order 4PN which is still negligible in the transformation to ADM
type coordinates.

The relative shift $\xi_{12}^i$ linking SH and ADM
coordinates, $x^{i}_{\rm ADM}=x^{i}_{\rm SH}+\xi_{12}^i$, is
given in~\cite{BI03CM} as\footnote{For simplicity we use the same notation
$\xi_{12}^i$ as for the shift between the SH and MH coordinates.}
\begin{eqnarray}\label{31'}
\xi_{12}^i &=& \frac{G\,m}{c^4}\Biggl\{\left[\frac{\dot{r}^2\,\nu}{8}
  - \frac{5\,\nu\,v^2}{8}+\frac{G\,m}{r}\left( -\frac{1}{4}- 3\,\nu
  \right) \right] n^i +\frac{9\,\dot{r}\,\nu}{4} v^i\Biggr\}
\nonumber\\ &+&
\frac{G\,m}{c^6}\Biggl\{\left[-\frac{\dot{r}^4\,\nu}{16} +
  \frac{5\,\dot{r}^4\,\nu^2}{16} + \frac{5\,\dot{r}^2\,\nu\,v^2}{16} -
  \frac{15\,\dot{r}^2\,\nu^2\,v^2}{16} - \frac{\nu\,v^4}{2} +
  \frac{11\,\nu^2\,v^4}{8} \right.\nonumber\\&&\qquad~ +
  \left.\frac{G\,m}{r}\left( \frac{161\,\dot{r}^2\,\nu}{48} -
  \frac{5\,\dot{r}^2\,\nu^2}{2} - \frac{451\,\nu\,v^2}{48} -
  \frac{3\,\nu^2\,v^2}{8} \right) \right.\nonumber\\&& \qquad~ +
  \left. \frac{G^2\,m^2}{r^2} \left( \frac{2773\,\nu}{280} +
  \frac{21\,\nu\,\pi^2}{32} - \frac{22\,\nu\,}{3}\ln
  \Big(\frac{r}{r'_0}\Big) \right)\right] n^i
\nonumber\nonumber\\&&\quad +\left[-\frac{5\,\dot{r}^3\,\nu}{12} +
  \frac{29\,\dot{r}^3\,\nu^2}{24} + \frac{17\,\dot{r}\,\nu\,v^2}{8} -
  \frac{21\,\dot{r}\,\nu^2\,v^2}{4} + \frac{G\,m}{r}\left(
  \frac{43\,\dot{r}\,\nu}{3} + 5\,\dot{r}\,\nu^2 \right)\right]
v^i\Biggr\}\,,
\end{eqnarray}
from which we deduce, applying Eqs.~\eqref{transf}, the transformation
of variables necessary to compute the ADM energy flux:
\begin{subequations}\label{A2ADM}\begin{align}
\delta_\xi r &= \frac{G \,m}{c^4} \left\{v^2\left( \frac{5}{8}\nu
\right)+\dot{r}^2 \left(-\frac{19}{8}\nu \right)+\frac{G \,m}{r}
\left(\frac{1}{4}+3\nu \right)\right\}\nonumber\\& +\frac{G\, m \nu
}{c^6} \left\{v^4 \left(\frac{1}{2}-\frac{11}{8} \nu \right)+\dot{r}^2
v^2 \left(-\frac{39}{16}+\frac{99}{16} \nu
\right)\right.\nonumber\\&\qquad +\dot{r}^4
\left(\frac{23}{48}-\frac{73}{48} \nu \right)+\frac{G\, m}{r} v^2
\left(\frac{451}{48}+\frac{3}{8} \nu \right)\nonumber\\&\qquad
\left.+\frac{G\, m}{r} \dot{r}^2 \left(-\frac{283}{16}-\frac{5}{2} \nu
\right)+\frac{G^2\,m^2}{r^2} \left(-\frac{2773}{280}+\frac{22}{3}
\ln\left(\frac{r}{r'_0}\right)-\frac{21}{32} \pi
^2\right)\right\}\,,\\
\delta_\xi \dot{r} &= \frac{G\, m}{c^4 r}\dot{r} \left\{ v^2
\left(-\frac{19}{4}\nu \right)+\dot{r}^2 \left(\frac{19}{4}\nu
\right)+\frac{G\, m}{r} \left(-\frac{1}{4}+\frac{1}{2}\nu
\right)\right\}\nonumber\\& +\frac{G\, m \nu }{c^6 r}\dot{r}\left\{ v^4
\left(-\frac{39}{8}+\frac{99}{8} \nu \right)+\dot{r}^2 v^2
\left(\frac{163}{24}-\frac{443}{24} \nu \right)+\dot{r}^4
\left(-\frac{23}{12}+\frac{73}{12} \nu
\right)\right.\nonumber\\&\qquad +\frac{G\, m}{r} v^2
\left(-\frac{1603}{48}-\frac{17}{4} \nu \right)+\frac{G\, m}{r}
\dot{r}^2 \left(\frac{1777}{48}+\frac{131}{24} \nu
\right)\nonumber\\&\qquad \left.+\frac{G^2\,m^2}{r^2}\left(\frac{3121}{105}-\frac{44}{3}
\ln\left(\frac{r}{r'_0}\right)+\frac{21}{16} \pi ^2-\frac{11}{4} \nu
\right)\right\}\,,\\
\delta_\xi v^2 &= \frac{G\, m}{c^4 r} \left\{v^4 \left(-\frac{13}{4}\nu
\right)+\dot{r}^2 v^2\left( \frac{5}{2}\nu \right)+\dot{r}^4
\left(\frac{3}{4}\nu \right)\right.\nonumber\\&\qquad \left.+\frac{G\,
m}{r} v^2 \left(\frac{1}{2}+\frac{21}{2}\nu \right)+\frac{G\, m}{r}
\dot{r}^2 \left(-1-\frac{19}{2}\nu \right)\right\}\nonumber\\&
+\frac{G\, m \nu }{c^6 r} \left\{v^6 \left(-\frac{13}{4}+\frac{31}{4}
\nu \right)+\dot{r}^2 v^4 \left(\frac{31}{8}-\frac{75}{8} \nu
\right)+\dot{r}^4 v^2 \left(-\frac{3}{2} \nu \right)+\dot{r}^6
\left(-\frac{5}{8}+\frac{25}{8} \nu \right)\right.\nonumber\\&\qquad
+\frac{G\, m}{r} v^4 \left(-\frac{9}{8}-\frac{25}{4} \nu \right)+\frac{G\,
m}{r} \dot{r}^2 v^2 \left(-\frac{131}{8}+\frac{121}{4} \nu
\right)+\frac{G\, m}{r} \dot{r}^4 \left(\frac{99}{4}-\frac{259}{12} \nu
\right)\nonumber\\&\qquad +\frac{G^2\,m^2}{r^2} v^2
\left(-\frac{3839}{420}+\frac{44}{3}\ln\left(\frac{r}{r'_0}\right)-\frac{21}{16}\pi
^2+\nu \right)\nonumber\\&\qquad \left.+\frac{G^2\,m^2}{r^2} \dot{r}^2
\left(\frac{28807}{420}-44\ln\left(\frac{r}{r'_0}\right)+\frac{63}{16}
\pi ^2-\frac{13}{2} \nu \right)\right\}\,.
\end{align}
\end{subequations}
The above equations provide the 3PN generalization of Eq.~(4.6)
of~\cite{GI97}. They also incorporate the corrected transformation
between ADM and harmonic coordinates at 2PN, as given in~\cite{DGI04}.

Using the latter expressions, one finds that the SH energy flux is
changed by corrections at 2PN and 3PN relative orders given by (using
a notation similar to that introduced above -- \textit{i.e.} in which
the variables $r$, ${\dot r}^2$ and $v^2$ are considered as dummy
variables\footnote{${\cal F}_{\rm ADM}={\cal F}_{\rm SH} +\delta_\xi{\cal F}.$})
\begin{align} \label{ShtoADM}
\delta_\xi\mathcal{F} =& -\frac{G^4\,m^5 \nu
^2}{c^9r^5}\left\{\,\frac{184}{5} v^4 \nu -\frac{736}{5} \dot{r}^2 v^2
\nu \right.\nonumber\\&\qquad \left.+\frac{G\,m}{r} v^2
\left(\frac{16}{5}+\frac{48}{5} \nu \right) +\frac{320}{3} \dot{r}^4
\nu +\frac{G\,m}{r} \dot{r}^2 \left(-\frac{12}{5}-\frac{56}{15} \nu
\right) \right\}\nonumber\\& -\frac{G^4\,m^5 \nu ^2}{c^{11}r^5}
\left\{v^6 \left(\frac{5886}{35} \nu -\frac{1616}{7} \nu ^2\right)
+\,\dot{r}^2 v^4 \left(-\frac{129866}{105} \nu +\frac{21598}{15} \nu
^2\right) \right.\nonumber\\&\qquad +\frac{G \,m}{r} v^4
\left(-\frac{22798}{105} \nu +\frac{7528}{35} \nu ^2\right) +
\dot{r}^4 v^2 \left(\frac{689434}{315} \nu -\frac{714608}{315} \nu
^2\right) \nonumber\\&\qquad +\frac{G \,m}{r} \dot{r}^2 v^2
\left(-\frac{936}{35}+\frac{16103}{21} \nu -\frac{14086}{21} \nu
^2\right)\nonumber\\&\qquad +\frac{G^2\,m^2}{r^2} v^2
\left(-\frac{272}{7}+\left[-\frac{31856}{75}-\frac{42}{5} \pi
^2\right] \nu +\frac{96}{35} \nu ^2\right)\nonumber\\&\qquad
+\dot{r}^6 \left(-\frac{116138}{105} \nu +\frac{110986}{105}
\nu^2\right)\nonumber\\&\qquad +\frac{G \,m}{r} \dot{r}^4
\left(\frac{328}{15}-\frac{198097}{315} \nu +\frac{143924}{315} \nu
^2\right) \nonumber\\&\qquad +\frac{G^2\,m^2}{r^2} \dot{r}^2
\left(\frac{1612}{35}+\left(\frac{673544}{1575}+\frac{28}{5} \pi
^2\right) \nu +\frac{828}{35} \nu ^2\right) \nonumber\\&\qquad
+\frac{G^3 \,m^3}{r^3} \left(\frac{16}{35}+\frac{128}{35} \nu
-\frac{768}{35} \nu ^2\right)\nonumber\\&\qquad \left.+
\frac{G^2\,m^2}{r^2} \nu\left( \frac{1408}{15} v^2 -\frac{2816}{45}
\dot{r}^2 \right) \ln \left(\frac{r}{r'_0}\right)\right\}\,.
\end{align}
The examination of the coefficient of $\ln(r/r_0')$, given by the two
last terms 
of~\eqref{ShtoADM}, reveals that
this coefficient is the same as in the contact transformation from SH
to MH, given by~\eqref{EF-ShtoMh}. Therefore the contact
transformation from SH to ADM exactly cancels out the logarithms of SH
coordinates, and the final flux in ADM coordinates is free of $\ln
r_0'$. This is consistent with the general understanding that the $\ln
r_0'$ is a feature of a particular harmonic coordinate system and that
ADM coordinates do not yield complications associated with such
logarithms (the cancellation of the $\ln r_0'$ terms usually provides
a useful internal check of the computations).
%
\section{The generalized quasi-Keplerian representation}\label{QKrepr}
%
Before we discuss the calculation of the orbital average of the energy
flux in Sect.~\ref{avgE}, we must summarize the 3PN generalized
quasi-Keplerian representation of the binary motion recently obtained
by Memmesheimer, Gopakumar and Sch\"afer~\cite{MGS04}. Indeed, the
main application of the present computation is the evolution of the
orbital elements under GW radiation reaction to 3PN order. This
requires one to average over an orbit the instantaneous expressions
for the energy flux obtained in Sect.~\ref{EF-Sharm}. Averaging over
an orbit is most conveniently accomplished by the use of an explicit
solution of the equations of motion. The generalized quasi-Keplerian
(QK) representation of the motion at 3PN order~\cite{MGS04}
constitutes an essential input for the computations to follow.

The QK representation was introduced by Damour and
Deruelle~\cite{DD85} to discuss the problem of binary pulsar timing at
1PN order, where relativistic periastron precession first appears and
complicates the simpler Keplerian picture. This elegant formulation
also played an important  role in our computation of the hereditary terms
in Paper~I where we provided a summary of it. The 2PN extension of
this work in the ADM coordinates was next given in
Refs.~\cite{DS88,SW93,Wex95} and we shall now use the 3PN
parametrization in ADM and MH coordinates~\cite{MGS04}.

The radial motion is given in parametric form as\footnote{For
convenience in this paper we adapt somewhat the notation with respect
to the one in Ref.~\cite{MGS04}.}
\begin{subequations} \label{3PNQKR}
\begin{align}
 \label{3PNQKR-r}
r =& a_r \left( 1 -e_r\,\cos u \right)\,,\\
 \label{3PNQKR-t}
\ell =& u -e_t\,\sin u + f_t\,\sin V + g_t\,(V -u) + i_t\,\sin 2 V +
h_t\,\sin 3 V \,,
\end{align}\end{subequations}
while the corresponding angular motion is
\begin{align}\label{3PNQKR-phi}
\frac{\phi - \phi_\mathrm{P}}{K} =& V + f_\phi\,\sin 2V + g_\phi\,
\sin 3V + i_\phi\,\sin 4V + h_\phi\,\sin 5V \,.
\end{align}
The four angles $V$, $u$, $\ell$ and $\phi$ are respectively the true
anomaly, the eccentric anomaly, the mean anomaly and the orbital phase
($V$, $u$ and $\ell$ are measured from the periastron, and we denote
by $\phi_\mathrm{P}$ the value of $\phi$ at periastron). The mean
anomaly is proportional to the time elapsed since the instant
$t_\mathrm{P}$ of passage at periastron,
\begin{equation}\label{ell}
\ell=n\,\left(t-t_\mathrm{P}\right)\,,
\end{equation}
where $n=2\pi/P$ is the mean motion and $P$ is the orbital period. The
true anomaly $V$ is given by
\begin{equation}\label{3PNQKR-V}
V = 2 \arctan \biggl [ \biggl ( \frac{ 1 + e_{\phi}}{ 1 - e_{\phi}}
\biggr )^{1/2} \, \tan \frac{u}{2} \biggr ]\,.
\end{equation}
In the above $a_r$ represents the semi-major axis of the orbit, and
$e_r$, $e_t$, $e_\phi$ are three kinds of eccentricities, labelled
after the coordinates $t$, $r$ and $\phi$, and which differ from each
other starting at the 1PN order. The constant $K$ is linked with the
advance of periastron per orbital revolution, and is given by
$K=\Phi/(2\pi)$ where $\Phi$ is the angle of return to the
periastron. The notation $k\equiv K-1$ for the relativistic precession
is used in Paper~I and will also be useful here. The orbital elements
$f_t$, $f_\phi$, $g_t$, $g_\phi$, $\cdots$ parametrize the 2PN and 3PN
relativistic corrections, as will be clear from their expressions
below. [More precisely, $f_t$, $f_\phi$, $g_t$, $g_\phi$ are composed
of 2PN and 3PN terms, but $i_t$, $i_\phi$, $h_t$, $h_\phi$ start only
at 3PN order.]

Crucial to the formalism are the explicit formulas for all the orbital
elements and all the coefficients in Eqs.~\eqref{3PNQKR} above in
terms of the 3PN conserved orbital energy $E$ and angular momentum $J$
(divided by the binary's reduced mass). Recall that the construction
of a generalized quasi-Keplerian representation exploits the
fact that the radial equation -- which is given by Eq.~(2.1a) in
paper~I -- is a \textit{polynomial} in $1/r$ (of seventh degree at
3PN order). Therefore the presence of logarithmic terms in the SH
coordinates at 3PN order obstructs the construction of the QK
parametrization (at least by this method) and Ref.~\cite{MGS04} obtained it in
coordinates avoiding such logarithms, namely the MH and ADM
coordinates. In both ADM and MH coordinates the QK representation
takes the same form given by Eqs.~\eqref{3PNQKR}--~\eqref{3PNQKR-phi},
but of course the equations linking the orbital elements to 
$E$ and $J$ are different. These have been obtained as post-Newtonian
series up to 3PN order in Ref.~\cite{MGS04}. Since they form the basis
for our computation of the average energy flux, we provide the
complete relations here.

For convenience in the present paper we introduce a  PN
parameter which is directly linked to the energy $E$ and defined by
\begin{equation}\label{eps}
\varepsilon \equiv -\frac{2 E}{c^2}\,.
\end{equation}
[Recall that $E<0$ for gravitationally bound orbits.]
The equations to follow will then appear as PN expansions
in terms of $\varepsilon=\mathcal{O}(2)$. Also, we find useful to
define, in place of the angular momentum $J$,
\begin{equation}\label{j}
j \equiv -\frac{2 E J^2}{(G m)^2}\,.
\end{equation}
We have $j = -2E h^2$ in terms of the more usual definition $h\equiv
J/(G m)$. 
This parameter is at Newtonian order, $j=c^2 \varepsilon h^2={\cal{O}}\left(0\right)$.
The point is now to give all the orbital elements as PN
series in powers of $\varepsilon$ with coefficients depending on $j$
(and the dimensionless reduced mass ratio $\nu$). In ADM coordinates these are given
by~\cite{MGS04}
\begin{subequations}\begin{align}
n^\mathrm{ADM}=&\frac{\varepsilon^{3/2}\,c^3}{G\,m} \bigg\{
1+\frac{\varepsilon}{8}\, ( -15+\nu ) +\frac{\varepsilon^{2}}{128}
\biggl[ 555 +30\,\nu +11\,\nu^{2} + \frac{192}{j^{1/2}} ( -5+2\,\nu )
\biggr] \nonumber\\& + \frac{\varepsilon^{3}}{3072} \biggl[ -29385
-4995\,\nu-315\,\nu^{2}+135 \,\nu^{3} \nonumber\\&\quad -
\frac{16}{j^{3/2}} \bigg(
10080+123\,\nu\,\pi^{2}-13952\,\nu+1440\,\nu^{2}\bigg) +
\frac{5760}{j^{1/2}} (17 -9\,\nu+2\,\nu^{2} ) \biggr] \bigg\} \,, \\
K^\mathrm{ADM} =& 1+\frac{3\varepsilon}{j}+ \frac{\varepsilon^{2}}{4}
\biggl [ \frac{3}{j} ( -5+2\,\nu ) + \frac{15}{j^2} ( 7 -2\,\nu )
\biggr] +\frac{\varepsilon^{3}}{128} \biggl[ \frac{24}{j} ( 5 -5\nu+
4\nu^2) \nonumber\\&\quad - \frac{1}{j^2} \biggl( 10080
-13952\,\nu+123 \,\nu\,\pi^{2}+1440\,\nu^{2} \biggr)\nonumber\\&\quad
\quad+ \frac{5}{j^3} \biggl(7392-8000\,\nu+ 123\,\nu\,\pi^{2} +
336\,\nu^{2} \biggr) \biggr]\,, \\ a_r^\mathrm{ADM} =&
\frac{G\,m}{\varepsilon\,c^2}\bigg\{ 1+\frac{\varepsilon}{4} ( -7+\nu
) + \frac{\varepsilon^{2}}{16}\,\bigg[ 1+10\,\nu+{\nu}^{2}
+\frac{1}{j} (-68+44\,\nu) \bigg] \nonumber\\& + \frac
{\varepsilon^{3}}{192}\, \biggl[ 3-9\,\nu-6\,{\nu}^{2}
+3\,{\nu}^{3}+\frac{1}{j} \biggl( 864+ ( -3\,{\pi}^{2}-2212 )
\nu+432\,{\nu}^{2}\biggr)\nonumber\\ &\quad + \frac{1} {j^2} \biggl(
-6432+ ( 13488-240\,{\pi}^{2} ) \nu -768\,{\nu}^{2}\biggr) \biggr]
\bigg\}\,, \\ e_r^\mathrm{ADM} =& \Biggl[1 -j + \frac{\varepsilon}{4}
\biggl\{ 24 -4 \,\nu+5\,j (-3+ \nu ) \biggr\}\nonumber\\ & +
\frac{\varepsilon^2}{8} \biggl\{ 52+2\,\nu+2\,{\nu}^{2} -j (
80-55\,\nu+4\,{\nu}^{2} ) -\frac {8}{j} (-17+11\, \nu )\biggr\}
\nonumber\\& + \frac{\varepsilon^3}{192} \biggl\{
-768-6\,\nu\,{\pi}^{2} -344\,\nu-216\,{\nu}^{2} + 3 j\,\bigg(
-1488+1556\,\nu -319\,\nu^{2} +4\,{\nu}^{3} \bigg) \nonumber\\&\quad
-\frac{4}{j}\,\bigg( 588-8212\,\nu+177\,\nu
\,{\pi}^{2}+480\,{\nu}^{2}\bigg) +\frac{192}{j^2} \biggl ( 134-281
\,\nu+5\,\nu\,{\pi}^{2}+16\,{\nu}^{2} \biggr ) \biggr
\}\Biggr]^{1/2}\,, \\ e_t^\mathrm{ADM} =&\Biggl[1-j+
\frac{\varepsilon}{4}\, \bigg\{ -8+8\,\nu -j ( -17+7\,\nu ) \bigg\}
\nonumber\\& + \frac{\varepsilon^{2}}{8} \bigg\{8+4\,\nu
+20\,{\nu}^{2} - j( 112-47\,\nu +16\,\nu^{2}) -24\,j^{1/2}\, (
-5+2\,\nu )
\nonumber\\ &\quad+\frac{4}{j} (17 - 11\,\nu ) -\frac{24}{j^{1/2}} \, ( 5
-2\,\nu ) \bigg\}\nonumber\\ &\quad + \frac{\varepsilon^{3}}{192} \bigg\{
24\, ( -2+5\,\nu ) (-23+10\,\nu+ 4\,\nu^{2} ) -15\,j \biggl
(-528+200\,\nu-77\,\nu^{2} + 24\,\nu^{3} \biggr) \nonumber\\&\quad
-72\,j^{1/2}( 265-193\,\nu +46\,\nu^{2} ) - \frac{2}{j} \bigg( 6732
+117\,\nu\,{\pi }^{2}-12508\,\nu+2004\,{\nu}^{2}\bigg)
\nonumber\\&\quad + \frac{2}{j^{1/2}} \bigg(
16380-19964\,\nu+123\,\nu\,\pi^{2}+3240\,\nu^{2} \bigg)
\nonumber\\&\quad -\frac{2}{j^{3/2}} \bigg(
10080+123\,\nu\,\pi^{2}-13952\,\nu+1440\,\nu^{2} \bigg) +
\frac{96}{j^2} \bigg( 134 -281\,\nu+5\,\nu\,{\pi }^{2}+16\,{\nu}^{2}
\bigg) \bigg\}\Biggr]^{1/2}\,, \\ e_{\phi}^\mathrm{ADM}=& \Biggl[1 - j
+ \frac{\varepsilon}{4} \bigg\{ 24+ j (-15+\nu ) \bigg\}\nonumber\\&
\left.+\frac{\varepsilon^{2}}{16} \bigg\{ -32+ 176\,\nu+18\,\nu^{2} - j(
160-30 \,\nu + 3\,{\nu}^{2} ) + \frac{1}{j}\, ( 408 - 232\,\nu -
15\,\nu^{2} )\right\}\nonumber\\ & + \frac{\varepsilon^{3}}{384} \bigg\{
-16032+2764\,
\nu+3\,\nu\,\pi^{2}+4536\,\nu^{2}+234\,\nu^{3}\nonumber\\& -36\,j
\biggl( 248 -80\,\nu +13 \,\nu^{2}+\nu^{3} \biggr) - \frac{6}{j}
\bigg( 2456 - 26860\,\nu+581\,\nu\,\pi^{2}+2689\,\nu^{2}+10\,\nu^{3}
\bigg) \nonumber\\&\quad + \frac{3}{j^2} \biggl( 27776
-65436\,\nu+1325\,\nu\,\pi^{2}+3440\,\nu^{2}-70\,\nu^{3}
\biggr)\biggr\}\Biggr]^{1/2}\,,\\ f_t^\mathrm{ADM} =&
-\frac{\varepsilon^2}{8 j^{1/2}} \biggl\{ (4 + \nu)\,\nu \, \sqrt{1-j}
\biggr\} \nonumber\\+&\frac{\varepsilon^{3}}{192} \bigg\{
\frac{1}{j^{3/2}\sqrt{1 - j}} \bigg( 1728-4148\, \nu
+3\,\nu\,\pi^{2}+600\,\nu^{2}+33\,\nu^{3}\bigg) \nonumber\\+& 3\,
\frac{j^{1/2}}{\sqrt{ 1-j}} \nu\, (-64-4\,\nu + 23\,{\nu}^{2} )
\nonumber\\ +&
\frac{1}{\sqrt{ j\,(1 -j )}} \biggl ( -1728 + 4232\,\nu-3\,\nu\,{\pi
}^{2} -627\,{\nu}^{2}-105\,{\nu}^{3} \biggr ) \bigg\}\,, \\
g_t^\mathrm{ADM} =&\frac{3\,\varepsilon^{2}}{2}\,\biggl(\frac{5
-2\,\nu}{j^{1/2}} \biggr) \nonumber\\& +\frac{\varepsilon^{3}}{192}
\biggl\{ \frac{1}{j^{3/2}} \bigg( 10080+123\,\nu\,\pi^{2}-13952\,\nu
+1440\,{\nu}^{2} \bigg) + \frac{1}{j^{1/2}} ( -3420
+1980\,\nu-648\,{\nu}^{2} ) \biggr\}\,, \\ i_t^\mathrm{ADM}
=&\frac{\varepsilon^{3}}{32}\,\nu \frac{1 - j}{j^{3/2}} (23+12\,\nu+
6\,\nu^{2} )\,, \\ h_t^\mathrm{ADM} =& \frac{13\,\varepsilon^{3}}{192}
\nu^3 \biggl(\frac{ 1 - j }{j}\biggr)^{3/2}\,, \\ f_\phi^\mathrm{ADM}
=& \frac{\varepsilon^{2}}{8} \,\frac{1-j}{j^2}\, \nu \, (1 -3\,\nu)
\nonumber\\& +\frac{\varepsilon^{3}}{256} \bigg\{ \frac{4\,\nu}{j} (
-11-40\,\nu+24\,\nu^{2} )\nonumber\\&\quad + \frac{1}{j^2} \biggl(
-256 +1192\,\nu-49\,\nu\,\pi^{2} +336\,\nu^{2} -80\,\nu^{3}
\biggr)\nonumber\\&\quad + \frac{1}{j^3}
\biggl(256+49\,\nu\,\pi^{2}-1076\,\nu-384\,\nu^{2}-40\,\nu^{3} \biggr)
\bigg\}\,, \\ g_\phi^\mathrm{ADM} =& -\frac{3\varepsilon^{2}}{32}
\frac{\nu^2}{j^2} (1 -j)^{3/2} \nonumber\\&
+\frac{\varepsilon^3}{768}\sqrt{1 -j}\, \bigg\{ -\frac{3}{j}\,\nu^2\,
( 9-26 \,\nu ) - \frac{1}{j^2} \,\nu \biggl(
220+3\,\pi^{2}+312\,\nu+150\,\nu^{2} \biggr)\nonumber\\&\quad +
\frac{1}{j^3}\,\nu ( 220+3\,{\pi }^{2}+96\,\nu+45\,{\nu}^{2} )
\bigg\}\,, \\ i_\phi^\mathrm{ADM} =& \frac{\varepsilon^{3}}{128}
\,\frac{(1 -j)^2}{j^3}\,\nu ( 5+28\,\nu+10\,\nu^{2})\,, \\
h_\phi^\mathrm{ADM} =& \frac{5\,\varepsilon^3}{256}\,
\frac{\nu^3}{j^3} \,(1 -j)^{5/2} \,.
\end{align}\end{subequations}

The latter expressions are specific to the ADM coordinates and we want
to give now the corresponding expressions in MH coordinates. However
we recall first an important point related to the use of gauge
invariant variables in the elliptical orbit case as stressed by
Ref.~\cite{MGS04}. Indeed Damour and Sch\"afer~\cite{DS88} showed that
the functional form of $n$ and $K=\Phi/(2\pi)$ as functions of gauge
invariant variables like $\varepsilon$ and $j$ are identical in
different coordinate systems. Hence the expressions in MH coordinates
of these two parameters are the same as in ADM coordinates,
\begin{subequations}\label{nK}\begin{align}
n \equiv n^\mathrm{MH} =& n^\mathrm{ADM}\,, \\ K \equiv K^\mathrm{MH}
=& K^\mathrm{ADM}\,.
\end{align}\end{subequations}
This prompted Ref.~\cite{MGS04} to suggest the use of $n$ and $k=K-1$
as two gauge invariant variables in the general orbit
case.\footnote{Actually Ref.~\cite{MGS04} used
$x_\mathrm{MGS}=(G\,m\,n/c^3)^{2/3}$ together with $k'=k/3$.} In the
present work we propose to use a variant of the former
variables. Namely, instead of working with the mean motion $n$ we
shall systematically use the orbital frequency $\omega = K\,n$ as
defined in a general context in Sect.~II.A of Paper~I, and define as a
gauge invariant post-Newtonian parameter
\begin{equation}\label{x}
x=\left(\frac{G\,m\,\omega}{c^3}\right)^{2/3}\,.
\end{equation}
This choice constitutes the obvious generalization of the gauge
invariant variable $x$ used in the circular orbit case and will thus
facilitate the straightforward reading out and check of the circular
orbit limit. The parameter $x$ is related to the energy and angular
momentum variables $\varepsilon$ and $j$ up to 3PN order by 
\begin{eqnarray}\label{xepsj}
x&=&{\varepsilon \left\{1 +\varepsilon
\left(-\frac{5}{4}+\frac{1}{12}\nu
+2\frac{1}{j}\right)\right.}\nonumber\\&& {+\varepsilon
^2\left(\frac{5}{2}+\frac{5}{24}\nu +\frac{1}{18}\nu
^2+\frac{1}{j^{1/2}}(-5+2\nu
)+\frac{1}{j}\left(-5+\frac{7}{6}\nu
\right)+\frac{1}{j^2}\left(\frac{33}{2}-5\nu
\right)\right)}\nonumber\\&& {+\varepsilon
^3\left(-\frac{235}{48}-\frac{25}{24}\nu -\frac{25}{576}\nu
^2+\frac{35}{1296}\nu
^3+\frac{1}{j}\left(\frac{35}{4}-\frac{5}{3}\nu +\frac{25}{36}\nu
^2\right)\right.}\nonumber\\&&
{+\frac{1}{j^{1/2}}\left(\frac{145}{8}-\frac{235}{24}\nu
+\frac{29}{12}\nu
^2\right)+\frac{1}{j^{3/2}}\left(-45+\left(\frac{472}{9}-\frac{41}{96}\pi
^2\right)\nu -5\nu ^2\right)}\nonumber\\&&
{\left.\left.+\frac{1}{j^2}\left(-\frac{565}{8}+\left(\frac{1903}{24}-\frac{41}{64}\pi
^2\right)\nu -\frac{95}{12}\nu
^2\right)+\frac{1}{j^3}\left(\frac{529}{3}+\left(-\frac{610}{3}+\frac{205}{64}\pi
^2\right)\nu +\frac{35}{4}\nu ^2\right)\right)\right\}}.\nonumber\\&&
\end{eqnarray}
The other orbital elements are not gauge invariant and therefore their
expressions in MH coordinates differ at 2PN and 3PN orders from those
in ADM coordinates. We conclude by giving here all the needed
differences~\cite{MGS04}, 
\begin{subequations}
\begin{align}
a^{\rm MH}_r-a^{\rm ADM}_r =&
{G\,m\,\varepsilon\left(-\frac{5}{8}\nu
+\frac{1}{j}\left(\frac{1}{4}+\frac{17}{4}\nu
\right)\right)}\nonumber\\& +{G\,m\,\varepsilon
^2\left(\frac{1}{32}\nu +\frac{1}{32}\nu
^2+\frac{1}{j}\left(-\frac{1}{2}+\left(-\frac{11499}{560}+\frac{21}{32}\pi
^2\right)\nu +\frac{19}{4}\nu ^2\right)\right.}\nonumber\\&
{\left.+\frac{1}{j^2}\left(\frac{3}{2}+\left(\frac{14501}{420}-\frac{21}{16}\pi
^2\right)\nu -5\nu ^2\right)\right)}\,,
\\
e^{\rm MH}_r-e^{\rm ADM}_r =& {\frac{\varepsilon
^2}{\sqrt{1-j}}\left(\frac{1}{2}+\frac{73}{8}\nu -j\frac{5}{8}\nu
+\frac{1}{j}\left(-\frac{1}{2}-\frac{17}{2}\nu
\right)\right)}\nonumber\\& {+\frac{\varepsilon
^3}{\sqrt{1-j}}\left(\frac{13}{16}+\left(-\frac{5237}{1680}+\frac{21}{32}\pi
^2\right)\nu +\frac{19}{16}\nu ^2+j\left(-\frac{143}{64}\nu
+\frac{37}{64}\nu ^2\right)\right.}\nonumber\\&
{\left.+\frac{1}{j}\left(\frac{13}{8}+\left(\frac{3667}{56}-\frac{105}{32}\pi
^2\right)\nu -\frac{51}{4}\nu
^2\right)+\frac{1}{j^2}\left(-3+\left(-\frac{14501}{210}+\frac{21}{8}\pi
^2\right)\nu +10\nu ^2\right)\right)}\,,
\\
\label{diffet}
e^{\rm MH}_t-e^{\rm ADM}_t =& {\frac{\varepsilon
^2}{\sqrt{1-j}}\left(\frac{1}{4}+\frac{17}{4}\nu
\right)\left(1-\frac{1}{j}\right)+}\nonumber\\& {\frac{\varepsilon
^3}{\sqrt{1-j}}\left(-\frac{19}{32}-\frac{52}{3}\nu
+\frac{225}{32}\nu
^2+\frac{1}{j}\left(\frac{29}{16}+\left(\frac{79039}{1680}-\frac{21}{16}\pi
^2\right)\nu -\frac{201}{16}\nu ^2\right)\right.}\nonumber\\&
{\left.+\frac{1}{j^2}\left(-\frac{3}{2}+\left(-\frac{14501}{420}+\frac{21}{16}\pi
^2\right)\nu +5\nu ^2\right)\right)}\,,
\\
e^{\rm MH}_\phi-e^{\rm ADM}_\phi =& {\frac{\varepsilon
^2}{\sqrt{1-j}}\left(-\frac{1}{4}-\frac{71}{16}\nu +j\frac{1}{32}\nu
+\frac{1}{j}\left(\frac{1}{4}+\frac{141}{32}\nu
\right)\right)}\nonumber\\& {+\frac{\varepsilon
^3}{\sqrt{1-j}}\left(-\frac{13}{32}+\left(\frac{36511}{8960}-\frac{21}{128}\pi
^2\right)\nu -\frac{1723}{256}\nu ^2+j\left(\frac{17}{256}\nu
+\frac{33}{256}\nu ^2\right)\right.}\nonumber\\&
{\left.+\frac{1}{j}\left(-\frac{13}{16}+\left(-\frac{21817}{480}+\frac{147}{64}\pi
^2\right)\nu +\frac{169}{8}\nu
^2\right)+\frac{1}{j^2}\left(\frac{3}{2}+\left(\frac{621787}{13440}-\frac{273}{128}\pi
^2\right)\nu -\frac{1789}{128}\nu ^2\right)\right)}\,,
\\
f^{\rm MH}_{t}-f^{\rm ADM}_{t} =&{\varepsilon
^2\frac{19}{8}\frac{\sqrt{(1-j)}}{\sqrt{j}}\nu }+
{\frac{\varepsilon
^3}{\sqrt{j(1-j)}}\left(-1+\left(-\frac{296083}{6720}+\frac{21}{32}\pi
^2\right)\nu +\frac{989}{64}\nu ^2\right.}\nonumber\\&
{\left.+j\left(\frac{361}{64}\nu -\frac{171}{64}\nu
^2\right)+\frac{1}{j}\left(1+\left(\frac{276133}{6720}-\frac{21}{32}\pi
^2\right)\nu -\frac{799}{64}\nu ^2\right)\right)}\,,
\\
g^{\rm MH}_{t}-g^{\rm ADM}_{t} =&0\,,
\\
h^{\rm MH}_{t}-h^{\rm ADM}_{t} =&{-\frac{\varepsilon
^3}{192}(1-j)^{\frac{3}{2}}j^{-\frac{3}{2}}\nu (-23+73\nu )}\,,
\\
i^{\rm MH}_{t}-i^{\rm ADM}_{t} =&{-\frac{11}{32}\varepsilon
^3(1-j)j^{-\frac{3}{2}}\nu (-19+10\nu )}\,,
\\
f^{\rm MH}_{\phi}-f^{\rm ADM}_{\phi} =&{-\varepsilon
^2\left(\frac{1}{j}-\frac{1}{j^2}\right)\left(\frac{1}{8}+\frac{9}{4}\nu
\right)}+
{\frac{\varepsilon
^3}{j}\left(\frac{1}{32}+\frac{1045}{192}\nu -\frac{99}{32}\nu
^2
\right.}\nonumber\\&
+\frac{1}{j}\left(-\frac{5}{4}+\left(-\frac{139633}{3360}+\frac{21}{16}\pi
^2\right)\nu +\frac{117}{8}\nu ^2\right)
\nonumber\\&
{\left.+\frac{1}{j^2}\left(\frac{3}{2}+
\left(\frac{92307}{2240}-\frac{21}{16}\pi^2\right)\nu -\frac{351}{32}\nu ^2\right) \right)}\,,
\\
g^{\rm MH}_{\phi}-g^{\rm ADM}_{\phi} =&{\varepsilon
^2\frac{1}{32}\frac{{(1-j)^{3/2}}}{j^2}\nu }+
{{\varepsilon
^3}{\sqrt{1-j}}\left(\frac{1}{j}\left(\frac{7}{128}\nu
-\frac{5}{32}\nu
^2\right)+
\right.}\nonumber\\&
\frac{1}{j^2}\left(\left(-\frac{49709}{13440}+\frac{21}{128}\pi
^2\right)\nu +\frac{445}{128}\nu ^2\right)
{\left.+\frac{1}{j^3}\left(\left(\frac{100783}{26880}-\frac{21}{128}\pi
^2\right)\nu -\frac{847}{256}\nu ^2\right)\right)}\,,
\\
h^{\rm MH}_{\phi}-h^{\rm ADM}_{\phi} =&{-\frac{\varepsilon
^3}{256}(1-j)^{\frac{5}{2}}j^{-3}\nu (-1+5\nu )}\,,
\\
i^{\rm MH}_{\phi}-i^{\rm ADM}_{\phi} =&{-\frac{\varepsilon
^3}{384}(-1+j)^2j^{-3}\nu (-149+198\nu )}\,.
\end{align}
\end{subequations}
Finally we note that in the case of a circular orbit the angular
momentum variable, say $j_\odot$, is related to the constant of energy
$\varepsilon$ by the 3PN gauge-invariant expansion
\begin{equation}\label{jcirc}
j_\odot = 1+\frac{\varepsilon}{4}\, ( 9+\nu ) +
\frac{\varepsilon^2}{16}\, \biggl(81-32\nu+\nu^2\biggr) +
\frac{\varepsilon^3}{192} \biggl(2835-7699\nu+246\nu
\pi^2+96\nu^2+3\nu^3\biggr)\,,
\end{equation}
which is easily deduced using either MH or ADM coordinates.
This expression can be used to compute all the
orbital elements for circular orbits and we can check that
all of the eccentricities $e_r$, $e_t$ or $e_\phi$ are zero.
%
\section{Orbital average of the 3PN energy flux}
\label{avgE}
%
To average the energy flux over an orbit we will require the use of
the previous 3PN quasi-Keplerian representation of the
motion. Consequently, the averaging is only possible in MH or ADM
coordinates without the logarithms as discussed before. The average of
the (instantaneous part of the) energy flux is defined by
\begin{equation}
\langle\,\mathcal{F}_\mathrm{inst}\rangle = \frac{1}{P} \int^{P}_{0}
dt\,\mathcal{F}_\mathrm{inst}={1\over 2\,\pi} \int^{2\,\pi}_{0}
du\,{d\ell\over du}\,\mathcal{F}_\mathrm{inst}\,. \label{avg-integral}
\end{equation}
As we have seen the energy flux~\eqref{tot} is made of instantaneous
terms and hereditary (tail) terms. The hereditary terms have already
been computed and averaged in Paper~I.

Using the QK representation of the orbit discussed in
Sect.~\ref{QKrepr}, we can re-express the energy flux
$\mathcal{F}_\mathrm{inst}$ [or, more exactly,
$(d\ell/du)\,\mathcal{F}_\mathrm{inst}$], which is a function of its
natural variables $r$, $\dot{r}$ and $v^2$, as a function of the
frequency-related parameter $x$ defined by
Eqs.~\eqref{x}--\eqref{xepsj}, the ``time'' eccentricity $e_t$ and the
eccentric anomaly $u$.\footnote{Ref.~\cite{GI97} uses $G m/a_r$ and
$e_r$ while~\cite{DGI04} employs $G\,m\,n/c^3$ and $e_t$. We propose
the use of $x=(G\,m\,\omega/c^3)^{2/3}$ for reasons outlined in the
previous Section. The choice of $e_t$ rather than say $e_r$ is a
matter of convenience since it appears in the Kepler equation which
is directly dealt with when averaging over an orbit.} We note that in the expression of the energy
flux at the 3PN order there are some logarithmic terms of the type
$\ln (r/r_0)$ even in MH coordinates. Indeed, we recall that the MH
coordinates permit the removal of the log-terms $\ln (r/r'_0)$, where
$r'_0$ is the scale associated with Hadamard's self-field
regularization, but there are still the terms $\ln (r/r_0)$ which
involve the constant $r_0$ entering the definition of the multipole
moments for general sources. As a result we find that the general
structure of $\mathcal{F}_\mathrm{inst}$ (in MH or ADM coordinates) is
\begin{equation}
\label{QKE-3PN} {d\ell\over du}\,\mathcal{F}_\mathrm{inst} 
= \sum_{N=3}^{11}\left\{ \alpha_N(e_t)\,\frac{1}{(1- e_t\cos u)^{N}}
+\beta_N(e_t)\,\frac{\sin u}{(1- e_t\cos u)^{N}} +
\gamma_N(e_t)\,\frac{\ln(1-e_t \cos u)}{(1- e_t\cos u)^{N}}\right\}\,,
\end{equation}
where the coefficients $\alpha_N$, $\beta_N$, $\gamma_N$ so defined
are straightforwardly computed using the QK parametrization (they are
too long to be listed here). It is worth noting that the $\beta_N$'s
correspond to all the 2.5PN terms while the $\gamma_N$'s represent the
logarithmic terms at order 3PN. The dependence on the constant $\ln
r_0$ has been included into the coefficients $\alpha_N$'s. To compute
the average we have at our disposal some integration formulas. First
of all,
\begin{equation}\label{int0}
\frac{1}{2\pi}\int_0^{2\pi}\frac{\sin u }{(1-e \cos u)^{N}}du = 0\,,
\end{equation}
which shows that in the final result there will be no terms (of the
``instantaneous'' type) at 2.5PN order. The 2.5PN instantaneous
contribution is proportional to $\dot{r}$ and vanishes after averaging
since it includes only odd functions of $u$. Next, we have
\begin{equation}\label{int1}
\frac{1}{2\pi}\int_0^{2\pi}\frac{du}{(1-e \cos u)^{N}} =
\frac{(-)^{N-1}}{(N-1)!}
\left(\frac{d^{N-1}}{dy^{N-1}}\left[\frac{1}{\sqrt{y^2-e^2}}
\right]\right)_{y=1}\,,
\end{equation}
which can also be formulated with the help of the standard Legendre
polynomial $P_{N-1}$ as
\begin{equation}\label{25PN_Intg}
\frac{1}{2\pi}\int_0^{2\pi}\frac{du}{(1-e \cos u)^{N}} =
\frac{1}{(1-e^2)^{N/2}} P_{N-1} \left(\frac{1}{\sqrt{1-e^2}}\right)\,.
\end{equation}
Finally for the log-terms we have a  less trivial formula but
which takes a structure similar as in Eq.~\eqref{int1}, namely
\begin{equation}\label{int2}
\frac{1}{2\pi}\int_0^{2\pi}\frac{\ln(1-e \cos u)}{(1-e \cos
u)^{N}}\,du = \frac{(-)^{N-1}}{(N-1)!}
\left(\frac{d^{N-1}Y(y,e)}{dy^{N-1}}\right)_{y=1}\,,
\end{equation}
in which 
\begin{equation}\label{Yy}Y(y,e)=\frac{1}{\sqrt{y^2-e^2}}\left\{\ln
\left[\frac{\sqrt{1-e^2}+1}{2}\right]+2\ln\left[1 +
\frac{\sqrt{1-e^2}-1}{y +\sqrt{y^2-e^2}}\right]\right\}\,.
\end{equation}
%
\subsection{Orbital average in MH coordinates}\label{avgE-MH}
%
The expression for the instantaneous energy flux in MH coordinates is
given by Eqs.~\eqref{EF}--\eqref{3PNEF-inst} together with the
modification~\eqref{EF-ShtoMh} for transforming to MH
coordinates. Implementing all the above integrations, the flux can be
averaged over an orbit to order 3PN extending the results
of~\cite{GI97} at 2PN\footnote{Results of \cite{GI97} are given in ADM coordinates.}. The
result is presented in the form
\begin{equation}\label{AvEMh}
\langle\,\mathcal{F}_\mathrm{inst}\rangle =\frac{32 c^5}{5
G}\,\nu^2\,x^5\,\biggl(\mathcal{I}_\mathrm{N}^\mathrm{MH} +
x\,\mathcal{I}_\mathrm{1PN}^\mathrm{MH} + x^2\,\mathcal{I}_{\rm
2PN}^\mathrm{MH} + x^3\,\mathcal{I}_\mathrm{3PN}^\mathrm{MH} \biggr)\,,
\end{equation}
where the ``instantaneous'' post-Newtonian pieces
$\mathcal{I}_{n\mathrm{PN}}^\mathrm{MH}$ depend on $\nu$ and the time
eccentricity $e_t$ in MH coordinates (note that $e_t\equiv
e_t^\mathrm{MH}$ here), and read\footnote{The Newtonian coefficient
$\mathcal{I}_\mathrm{N}^\mathrm{MH}$ is nothing but the Peters \&
Mathews~\cite{PM63} ``enhancement'' function of eccentricity
$f(e_t)\equiv (1+\frac{73}{24}e_t^2 +
\frac{37}{96}e_t^4)/(1-e_t^2)^{7/2}$, called that way because it
enhances the numerical value of the orbital decay of the binary pulsar
by gravitational radiation (\textit{viz} the orbital $\dot{P}$).}

\begin{subequations}
\label{AvEMha}
\begin{align}
\label{AvEMh0} \mathcal{I}_\mathrm{N}^\mathrm{MH} =&\frac{1}{(1-e_t^2)^{7/2}}{\left\{
1+\frac{73}{24}~e_t^2 + \frac{37}{96}~e_t^4\right\}}\,,\\
\label{AvEMh1} \mathcal{I}_\mathrm{1PN}^\mathrm{MH} =&
\frac{1}{(1-e_t^2)^{9/2}}{\left\{-\frac{1247}{336} -\frac{35}{12} \nu
  +e_t^2\left(\frac{10475}{672}-\frac{1081}{36} \nu
  \right)\right.}\nonumber\\&
  {\left.+e_t^4\left(\frac{10043}{384}-\frac{311}{12} \nu
  \right)+e_t^6\left(\frac{2179}{1792}-\frac{851}{576} \nu
  \right)\right\}}\,,\\
\label{AvEMh2} \mathcal{I}_\mathrm{2PN}^\mathrm{MH} =&\frac{1}{(1-e_t^2)^{11/2}}
{\left\{-\frac{203471}{9072}+\frac{12799}{504} \nu +\frac{65}{18} \nu
^2\right.}\nonumber\\&
{+e_t^2\left(-\frac{3807197}{18144}+\frac{116789}{2016} \nu
+\frac{5935}{54} \nu ^2\right)}\nonumber\\&
{+e_t^4\left(-\frac{268447}{24192}-\frac{2465027}{8064} \nu
+\frac{247805}{864} \nu ^2\right)}\nonumber\\&
{+e_t^6\left(\frac{1307105}{16128}-\frac{416945}{2688} \nu
+\frac{185305}{1728} \nu ^2\right)}\nonumber\\&
{+e_t^8\left(\frac{86567}{64512}-\frac{9769}{4608} \nu
+\frac{21275}{6912} \nu ^2\right)}\nonumber\\&
{+\sqrt{1-e_t^2}\left[\frac{35}{2}-7 \nu
+e_t^2\left(\frac{6425}{48}-\frac{1285}{24} \nu
\right)\right.}\nonumber\\&
{\left.\left.+e_t^4\left(\frac{5065}{64}-\frac{1013}{32} \nu
\right)+e_t^6\left(\frac{185}{96}-\frac{37}{48} \nu
\right)\right]\right\}}\,,\\
\label{AvEMh3} \mathcal{I}_\mathrm{3PN}^\mathrm{MH} =& \frac{1} {(1-e_t^2)^{13/2}}
{\left\{
\frac{2193295679}{9979200}+\left[\frac{8009293 }{54432}-\frac{41 \pi ^2  }{64}\right]\nu-\frac{209063 }{3024}\nu ^2-\frac{775}{324} \nu ^3
\right.}
\nonumber\\& {+e_t^2\left(
\frac{20506331429}{19958400}+\left[\frac{649801883}{272160}+\frac{4879 \pi ^2}{1536}\right]\nu-\frac{3008759 }{3024}\nu ^2-\frac{53696 }{243}\nu ^3
\right)}\nonumber\\& {+e_t^4\left(
-\frac{3611354071}{13305600}+\left[\frac{755536297 }{136080}-\frac{29971 \pi ^2 }{1024}\right]\nu-\frac{179375 }{576}\nu
   ^2-\frac{10816087 }{7776}\nu ^3
\right)}\nonumber\\& {+e_t^6\left(
\frac{4786812253}{26611200}+\left[\frac{1108811471 }{1451520}-\frac{84501 \pi ^2 }{4096}\right]\nu+\frac{87787969 }{48384}\nu
   ^2-\frac{983251 }{648}\nu ^3
\right)}\nonumber\\& {+e_t^8\left(
\frac{21505140101}{141926400}+\left[-\frac{32467919}{129024}-\frac{4059 \pi ^2 }{4096}\right]\nu+\frac{79938097 }{193536}\nu
   ^2-\frac{4586539 }{15552}\nu ^3
\right)}\nonumber\\&
{+e_t^{10}\left(
-\frac{8977637}{11354112}+\frac{9287 }{48384}\nu+\frac{8977 }{55296}\nu ^2-\frac{567617 }{124416}\nu ^3
\right)}\nonumber\\&
+\sqrt{1-e_t^2}\left[
-\frac{14047483}{151200}+\left[-\frac{165761 }{1008}+\frac{287 \pi ^2  }{192}\right]\nu+\frac{455 }{12}\nu ^2\right. \nonumber\\&
+e_t^2\left(
\frac{36863231}{100800}+\left[-\frac{14935421}{6048}+\frac{52685 \pi ^2 }{4608}\right]\nu+\frac{43559 }{72}\nu ^2
\right) \nonumber\\&+e_t^4\left(
\frac{759524951}{403200}+\left[-\frac{31082483 }{8064}+\frac{41533 \pi ^2  }{6144}\right]\nu+\frac{303985 }{288}\nu ^2
\right)\nonumber\\&
+e_t^6\left(
\frac{1399661203}{2419200}+\left[-\frac{40922933}{48384}+\frac{1517 \pi ^2}{9216}\right]\nu+\frac{73357 }{288}\nu ^2
\right)\nonumber\\&
+ \left.e_t^8\left(
\frac{185}{48}-\frac{1073  }{288}\nu+\frac{407 }{288}\nu ^2
\right)\right]
\nonumber\\& \left. + \left(\frac{1712}{105}+\frac{14552}{63}
e_t^2+\frac{553297}{1260} e_t^4+\frac{187357}{1260}
e_t^6+\frac{10593}{2240} e_t^8\right)
\ln\left[\frac{x}{x_0}\frac{1+\sqrt{1-e_t^2}}{2(1-e_t^2)}\right]\right\}\,.
\end{align}
\end{subequations}
For ease of presentation we have not put a label on
$e_t$ to indicate that it is the time eccentricity in MH coordinates. Of course, since $x$ is gauge invariant, no such label is
required on it. It is important to keep track of this fact when
comparing formulas in different gauges, as we will eventually do.

The last term in the 3PN coefficient
$\mathcal{I}_\mathrm{3PN}^\mathrm{MH}$ given by Eq.~\eqref{AvEMh3} is
proportional to some logarithm which directly arises from the
integration formula~\eqref{int2}--\eqref{Yy}. Inside the logarithm we
posed
\begin{equation}\label{x0}
x_0 \equiv \frac{G\,m}{c^2\,r_0}\,,
\end{equation}
exhibiting the dependence of the instantaneous part of the 3PN energy
flux upon the arbitrary constant length scale $r_0$. Only after
computing the complete energy flux can one discuss the structure of
the logarithmic term in the energy flux and the required cancellation
of the $\ln r_0$. Therefore we now add the hereditary contribution to
the 3PN flux which has been computed in Paper~I. From Eq.~(6.2) in
Paper~I we write the result as
\begin{equation}\label{hered1}
\langle\,\mathcal{F}_\mathrm{hered}\rangle =\frac{32 c^5}{5
G}\,\nu^2\,x^5\,\biggl(
x^{3/2}\,\mathcal{H}_\mathrm{1.5PN}^\mathrm{MH} +
x^{5/2}\,\mathcal{H}_\mathrm{2.5PN}^\mathrm{MH} +
x^3\,\mathcal{H}_\mathrm{3PN}^\mathrm{MH} \biggr)\,,
\end{equation}
where the ``hereditary'' post-Newtonian coefficients (starting at
1.5PN order) read
\begin{subequations}\label{hered2}\begin{align}
\mathcal{H}_\mathrm{1.5N}^\mathrm{MH} =&\, 4\pi\,\varphi(e_t)\,,\\
\mathcal{H}_{\rm 2.5PN}^\mathrm{MH} =&
-\frac{8191}{672}\,\pi\,\psi(e_t)
-\frac{583}{24}\nu\,\pi\,\zeta(e_t)\,,\\
\mathcal{H}_\mathrm{3PN}^\mathrm{MH} =&
-\frac{116761}{3675}\,\kappa(e_t) + \left[\frac{16}{3} \,\pi^2
-\frac{1712}{105}\,C -
\frac{1712}{105}\ln\left(\frac{4x^{3/2}}{x_0}\right)\right]\,F(e_t)\,.
\label{hered2log}
\end{align}
\end{subequations}
The function $F(e_t)$ in factor of the logarithm in the 3PN
coefficient does admit a closed analytic form which was determined in
paper~I as
\begin{equation}\label{Fet}
F(e_t) = \frac{1}{(1-e_t^2)^{13/2}}
\left[1+\frac{85}{6}e_t^2+\frac{5171}{192}e_t^4+\frac{1751}{192}e_t^6
+\frac{297}{1024}e_t^8\right]\,.
\end{equation}
On the other hand paper~I found that the four ``enhancement''
functions of eccentricity $\varphi(e_t)$, $\psi(e_t)$, $\zeta(e_t)$
and $\kappa(e_t)$ very likely do not admit  any analytic closed-form
expressions. Numerical plots of the four enhancement factors
$\varphi(e_t)$, $\psi(e_t)$, $\theta(e_t)$ and $\kappa(e_t)$ as
functions of eccentricity $e_t$ have been presented in Paper~I. The
coefficients in Eqs.~\eqref{hered2} have been introduced in such a way
that the circular orbit limit of all the functions $F(e_t)$ and
$\varphi(e_t)$, $\cdots$, $\kappa(e_t)$ is one.

Finally, the PN coefficients in the total averaged energy flux
$\mathcal{F}$ in MH coordinates are given by the sum of the
instantaneous and hereditary contributions, say
\begin{equation}\label{K}
\mathcal{K}_{n\mathrm{PN}}^\mathrm{MH}=\mathcal{I}_{n\mathrm{PN}}^\mathrm{MH}
+\mathcal{H}_{n\mathrm{PN}}^\mathrm{MH}\,.
\end{equation}
We notice that up to 2.5PN order there is a clean separation between
the instantaneous terms which are at even PN orders (recall that there
is no 2.5PN term in the averaged flux), and the hereditary terms which
appear at odd PN orders and are specifically due to tails
(\textit{i.e.} $\mathcal{H}_\mathrm{1.5PN}^\mathrm{MH}$ and
$\mathcal{H}_\mathrm{2.5PN}^\mathrm{MH}$). On the contrary, at 3PN
order -- and, indeed, at any higher PN order -- there is a mixture
of instantaneous and hereditary terms. The 3PN hereditary term
$\mathcal{H}_\mathrm{3PN}^\mathrm{MH}$ is due to the so-called GW
tails of tails (see Paper~I).

The analytical result~\eqref{Fet} is crucial for checking that the
arbitrary constant $x_0$ disappears from the final result, namely from
the 3PN coefficient $\mathcal{K}_\mathrm{3PN}^\mathrm{MH}$. Indeed we
immediately verify from comparing the last term in Eq.~\eqref{AvEMh3}
with Eq.~\eqref{hered2log} and the explicit expression~\eqref{Fet} of
$F(e_t)$ that $x_0$ cancels out from the sum of the instantaneous and
hereditary contributions, extending to non circular orbits this fact
which was already observed for the circular case in
Ref.~\cite{BIJ02}. Finally the complete 3PN coefficient (independent of
$x_0$) reads
\begin{align}\label{K3PN} 
\mathcal{K}_\mathrm{3PN}^\mathrm{MH} =& 
\frac{1} {(1-e_t^2)^{13/2}}
{\left\{
\frac{2193295679}{9979200}+\left[\frac{8009293 }{54432}-\frac{41 \pi ^2  }{64}\right]\nu-\frac{209063 }{3024}\nu ^2-\frac{775}{324} \nu ^3
\right.}
\nonumber\\& {+e_t^2\left(
\frac{20506331429}{19958400}+\left[\frac{649801883}{272160}+\frac{4879 \pi ^2}{1536}\right]\nu-\frac{3008759 }{3024}\nu ^2-\frac{53696 }{243}\nu ^3
\right)}\nonumber\\& {+e_t^4\left(
-\frac{3611354071}{13305600}+\left[\frac{755536297 }{136080}-\frac{29971 \pi ^2 }{1024}\right]\nu-\frac{179375 }{576}\nu
   ^2-\frac{10816087 }{7776}\nu ^3
\right)}\nonumber\\& {+e_t^6\left(
\frac{4786812253}{26611200}+\left[\frac{1108811471 }{1451520}-\frac{84501 \pi ^2 }{4096}\right]\nu+\frac{87787969 }{48384}\nu
   ^2-\frac{983251 }{648}\nu ^3
\right)}\nonumber\\& {+e_t^8\left(
\frac{21505140101}{141926400}+\left[-\frac{32467919}{129024}-\frac{4059 \pi ^2 }{4096}\right]\nu+\frac{79938097 }{193536}\nu
   ^2-\frac{4586539 }{15552}\nu ^3
\right)}\nonumber\\&
{+e_t^{10}\left(
-\frac{8977637}{11354112}+\frac{9287 }{48384}\nu+\frac{8977 }{55296}\nu ^2-\frac{567617 }{124416}\nu ^3
\right)}\nonumber\\&
+\sqrt{1-e_t^2}\left[
-\frac{14047483}{151200}+\left[-\frac{165761 }{1008}+\frac{287 \pi ^2  }{192}\right]\nu+\frac{455 }{12}\nu ^2\right. \nonumber\\&
+e_t^2\left(
\frac{36863231}{100800}+\left[-\frac{14935421}{6048}+\frac{52685 \pi ^2 }{4608}\right]\nu+\frac{43559 }{72}\nu ^2
\right) \nonumber\\&+e_t^4\left(
\frac{759524951}{403200}+\left[-\frac{31082483 }{8064}+\frac{41533 \pi ^2  }{6144}\right]\nu+\frac{303985 }{288}\nu ^2
\right)\nonumber\\&
+e_t^6\left(
\frac{1399661203}{2419200}+\left[-\frac{40922933}{48384}+\frac{1517 \pi ^2}{9216}\right]\nu+\frac{73357 }{288}\nu ^2
\right)\nonumber\\&
+ \left.e_t^8\left(
\frac{185}{48}-\frac{1073  }{288}\nu+\frac{407 }{288}\nu ^2
\right)\right]
\nonumber\\& 
+ \left(\frac{1712}{105}+\frac{14552}{63}
e_t^2+\frac{553297}{1260} e_t^4 +\frac{187357}{1260}
e_t^6+\frac{10593}{2240} e_t^8\right)\nonumber\\&\qquad\left.\times
\left[ - C +\frac{35}{107}\pi^2 - \frac{1}{2}\ln\left(16 x\right) +
\ln\left(\frac{1+\sqrt{1-e_t^2}}{2(1-e_t^2)}\right)\right]\right\}
-\frac{116761}{3675}\kappa(e_t)\,.
\end{align}
The 1.5PN and 2.5PN coefficients are only due to tails, thus
\begin{subequations}\label{Ktail}\begin{align}
\mathcal{K}_\mathrm{1.5N}^\mathrm{MH} =&\, 4\pi\,\varphi(e_t)\,,\\
\mathcal{K}_{\rm 2.5PN}^\mathrm{MH} =&
-\frac{8191}{672}\,\pi\,\psi(e_t)
-\frac{583}{24}\nu\,\pi\,\zeta(e_t)\,.
\end{align}
\end{subequations}
The Newtonian, 1PN and 2PN coefficients reduce to their instantaneous
contributions $\mathcal{I}_\mathrm{N}^\mathrm{MH}$,
$\mathcal{I}_\mathrm{1PN}^\mathrm{MH}$ and
$\mathcal{I}_\mathrm{2PN}^\mathrm{MH}$ already given in
Eqs.~\eqref{AvEMha}.

Since the enhancement functions $\varphi(e_t)$, $\psi(e_t)$,
$\zeta(e_t)$ and $\kappa(e_t)$ reduce to one in the circular case,
when $e_t=0$, the circular-orbit limit of the energy flux is
immediately deduced from inspection of Eqs.~\eqref{AvEMha}
and~\eqref{Ktail} as
\begin{align}\label{Flux_CL_x}
\langle \mathcal{F}\rangle_{\odot} =& \frac{32c^5}{5G}
x^5\,\nu^2\,\left\{1+x \left(-\frac{1247}{336}-\frac{35}{12}\nu\right)
+ 4\pi\,x^{3/2} \right.\nonumber\\& + x^2
\left(-\frac{44711}{9072}+\frac{9271}{504} \nu +\frac{65}{18}
\nu^2\right) + \pi\,x^{5/2} \left(-\frac{8191}{672} -\frac{583}{24}\nu
\right) \nonumber\\& + x^3
\left(\frac{6643739519}{69854400}-\frac{1712}{105}C+\frac{16}{3} \pi^2
- \frac{856}{105} \ln\left(16 x\right) \right.\nonumber\\&
\qquad\left.\left. +\left[-\frac{134543}{7776}+\frac{41}{48}
\pi^2\right] \nu -\frac{94403}{3024} \nu^2-\frac{775}{324} \nu
^3\right)\right\}\,.
\end{align}
This limiting case is in exact agreement with Eq.~(12.9)
of~\cite{BIJ02} (after taking into account the values of the ambiguity
parameters $\lambda=-\frac{1987}{3080}$ and
$\theta=-\frac{11831}{9240}$ computed in
Refs.~\cite{BDEI04,BDI04zeta,BDEI05}). Notice that the flux in the
circular-orbit limit~\eqref{Flux_CL_x} depends only on the parameter
$x$ and hence its expression becomes gauge invariant.
%
\subsection{Orbital average in ADM coordinates}\label{avgE-ADM}
%
We start from the expression for the instantaneous energy flux in ADM
coordinates as given by~\eqref{ShtoADM}, employ the appropriate 3PN QK
representation and follow the procedure for performing the average as
outlined in the previous Section. We find that the $\beta_N$'s and
$\gamma_N$'s in ADM coordinates [\textit{cf.} Eq.~\eqref{QKE-3PN}] are
exactly the same as in MH coordinates; the $\alpha_N$'s, however, are
different in general (except for $\alpha_{11}$). The result for the
average energy flux in ADM coordinates is of the form
\begin{equation}\label{AvEADM}
\langle\,\mathcal{F}_\mathrm{inst}\rangle =\frac{32 c^5}{5
G}\,\nu^2\,x^5\,\biggl(\mathcal{I}_\mathrm{N}^\mathrm{ADM} + x\,\mathcal{I}_{\rm
1PN}^\mathrm{ADM} + x^2\,\mathcal{I}_\mathrm{2PN}^\mathrm{ADM} + x^3\,\mathcal{I}_{\rm
3PN}^\mathrm{ADM} \biggr)\,,
\end{equation}
where the coefficients depend on the time eccentricity in ADM
coordinates (hence $e_t\equiv e_t^\mathrm{ADM}$ here) and on $\nu$,
and read
\begin{subequations}\label{AvEADMa}\begin{align}
\label{AvEADM0} \mathcal{I}_\mathrm{N}^\mathrm{ADM} =& \frac{1}{(1-e_t^2)^{7/2}}\left\{1
+\frac{73}{24}\,e_t^2+\frac{37}{96}\,e_t^4\right\}\,,\\
\label{AvEADM1} \mathcal{I}_\mathrm{1PN}^\mathrm{ADM} =&{\frac{1}{(1-e_t^2)^{9/2}}
\left\{-\frac{1247}{336}-\frac{35}{12} \nu
+e_t^2\left(\frac{10475}{672}-\frac{1081}{36} \nu
\right)\right.}\nonumber\\&
{\left.+e_t^4\left(\frac{10043}{384}-\frac{311}{12} \nu
\right)+e_t^6\left(\frac{2179}{1792}-\frac{851}{576} \nu
\right)\right\}}\,,\\
\label{AvEADM2} \mathcal{I}_\mathrm{2PN}^\mathrm{ADM} =& 
\frac{1}{(1-e_t^2)^{11/2}}\left\{-\frac{203471}{9072}
+\frac{12799}{504} \nu +\frac{65}{18}
\nu^2\right.\nonumber\\&+e_t^2\left(-\frac{3866543}{18144}+\frac{4691}{2016}
\nu +\frac{5935}{54} \nu ^2\right)\nonumber\\&
{+e_t^4\left(-\frac{369751}{24192}-\frac{3039083}{8064} \nu
+\frac{247805}{864} \nu ^2\right)}\nonumber\\&
{+e_t^6\left(\frac{1302443}{16128}-\frac{215077}{1344} \nu
+\frac{185305}{1728} \nu ^2\right)}\nonumber\\&
{+e_t^8\left(\frac{86567}{64512}-\frac{9769}{4608} \nu
+\frac{21275}{6912} \nu ^2\right)}\nonumber\\&
{+\sqrt{1-e_t^2}\left[\frac{35}{2}-7 \nu
+e_t^2\left(\frac{6425}{48}-\frac{1285}{24} \nu
\right)\right.}\nonumber\\&
{\left.\left.+e_t^4\left(\frac{5065}{64}-\frac{1013}{32} \nu
\right)+e_t^6\left(\frac{185}{96}-\frac{37}{48} \nu
\right)\right]\right\}}\,,\\
\label{AvEADM3} \mathcal{I}_\mathrm{3PN}^\mathrm{ADM} =&
 \frac{1}{(1-e_t^2)^{13/2}}
\left\{\frac{2193295679}{9979200}+\left[\frac{8009293 
   }{54432}-\frac{41 \pi ^2 }{64}\right]\nu-\frac{209063 }{3024}\nu
   ^2-\frac{775 }{324}\nu ^3
\right.\nonumber\\&
{+e_t^2\left(
\frac{2912411147}{2851200}+\left[\frac{249108317 
   }{108864}+\frac{31255}{1536}\pi^2\right]\nu-\frac{3525469
   }{6048}\nu ^2-\frac{53696 }{243}\nu ^3
\right)}
\nonumber\\& {+e_t^4\left(
-\frac{4520777971}{13305600}+\left[\frac{473750339 
   }{108864}-\frac{7459 \pi ^2 }{1024}\right]\nu+\frac{697997
   }{576}\nu ^2-\frac{10816087}{7776} \nu ^3
\right)}\nonumber\\& {+e_t^6\left(
\frac{3630046753}{26611200}+\left[-\frac{8775247}{145152}-\frac{78285 \pi ^2 }{4096}\right]\nu+\frac{31147213 }{12096}\nu ^2-\frac{983251
  }{648} \nu ^3
\right)}\nonumber\\& {+e_t^8\left(
\frac{21293656301}{141926400}+\left[-\frac{36646949}{129024}-\frac{4059 \pi ^2}{4096}\right]\nu+\frac{85830865 }{193536}\nu
   ^2-\frac{4586539 }{15552}\nu ^3
\right)}\nonumber\\&
{+e_t^{10}\left(-\frac{8977637}{11354112}+\frac{9287}{48384} \nu
+\frac{8977}{55296} \nu ^2-\frac{567617}{124416} \nu
^3\right)}\nonumber\\&
+\sqrt{1-e_t^2}\left[
-\frac{14047483}{151200}+\left[-\frac{165761 }{1008}+\frac{287 \pi ^2 }{192}\right]\nu+\frac{455 }{12}\nu ^2\right.\nonumber\\&
+e_t^2\left(\frac{36863231}{100800}+\left[-\frac{14935421}{6048}+\frac{52685 \pi ^2}{4608}\right]\nu+\frac{43559 }{72}\nu ^2\right) \nonumber\\&
+e_t^4\left(\frac{759524951}{403200}+\left[-\frac{31082483}{8064}+\frac{41533 \pi ^2}{6144}\right]\nu+\frac{303985 }{288}\nu ^2\right)\nonumber\\&
+e_t^6\left(\frac{1399661203}{2419200}+\left[-\frac{40922933 }{48384}+\frac{1517 \pi ^2}{9216}\right]\nu+\frac{73357 }{288}\nu ^2\right)\nonumber\\&
+\left.e_t^8\left(\frac{185}{48}-\frac{1073 }{288}\nu+\frac{407 }{288}\nu ^2\right)\right]\nonumber\\&
 \left. + \left(\frac{1712}{105}+\frac{14552}{63}
e_t^2+\frac{553297}{1260} e_t^4+\frac{187357}{1260}
e_t^6+\frac{10593}{2240} e_t^8\right)
\ln\left[\frac{x}{x_0}\frac{1+\sqrt{1-e_t^2}}{2(1-e_t^2)}\right]\right\}\,.
\end{align}
\end{subequations}
We recall that the Newtonian and 1PN orders are the same in MH and ADM
coordinates (the coefficients $\mathcal{I}_\mathrm{N}^\mathrm{ADM}$
and $\mathcal{I}_\mathrm{1PN}^\mathrm{ADM}$ agree with their MH
counterparts). On the other hand, adding up the hereditary
contribution~\eqref{hered1}--\eqref{hered2} [which is the same in MH
and ADM coordinates], we obtain the total 3PN coefficient
$\mathcal{K}_\mathrm{3PN}^\mathrm{ADM}$, analogous to Eq.~\eqref{K3PN}
but in ADM coordinates,
\begin{align}\label{K3PNadm} 
\mathcal{K}_\mathrm{3PN}^\mathrm{ADM} =& \frac{1}{(1-e_t^2)^{13/2}}
\left\{\frac{2193295679}{9979200}+\left[\frac{8009293 
   }{54432}-\frac{41 \pi ^2 }{64}\right]\nu-\frac{209063 }{3024}\nu
   ^2-\frac{775 }{324}\nu ^3
\right.\nonumber\\&
{+e_t^2\left(
\frac{2912411147}{2851200}+\left[\frac{249108317 
   }{108864}+\frac{31255}{1536}\pi^2\right]\nu-\frac{3525469
   }{6048}\nu ^2-\frac{53696 }{243}\nu ^3
\right)}
\nonumber\\& {+e_t^4\left(
-\frac{4520777971}{13305600}+\left[\frac{473750339 
   }{108864}-\frac{7459 \pi ^2 }{1024}\right]\nu+\frac{697997
   }{576}\nu ^2-\frac{10816087}{7776} \nu ^3
\right)}\nonumber\\& {+e_t^6\left(
\frac{3630046753}{26611200}+\left[-\frac{8775247}{145152}-\frac{78285 \pi ^2 }{4096}\right]\nu+\frac{31147213 }{12096}\nu ^2-\frac{983251
  }{648} \nu ^3
\right)}\nonumber\\& {+e_t^8\left(
\frac{21293656301}{141926400}+\left[-\frac{36646949}{129024}-\frac{4059 \pi ^2}{4096}\right]\nu+\frac{85830865 }{193536}\nu
   ^2-\frac{4586539 }{15552}\nu ^3
\right)}\nonumber\\&
{+e_t^{10}\left(-\frac{8977637}{11354112}+\frac{9287}{48384} \nu
+\frac{8977}{55296} \nu ^2-\frac{567617}{124416} \nu
^3\right)}\nonumber\\&
+\sqrt{1-e_t^2}\left[
-\frac{14047483}{151200}+\left[-\frac{165761 }{1008}+\frac{287 \pi ^2 }{192}\right]\nu+\frac{455 }{12}\nu ^2\right.\nonumber\\&
+e_t^2\left(\frac{36863231}{100800}+\left[-\frac{14935421}{6048}+\frac{52685 \pi ^2}{4608}\right]\nu+\frac{43559 }{72}\nu ^2\right) \nonumber\\&
+e_t^4\left(\frac{759524951}{403200}+\left[-\frac{31082483}{8064}+\frac{41533 \pi ^2}{6144}\right]\nu+\frac{303985 }{288}\nu ^2\right)\nonumber\\&
+e_t^6\left(\frac{1399661203}{2419200}+\left[-\frac{40922933 }{48384}+\frac{1517 \pi ^2}{9216}\right]\nu+\frac{73357 }{288}\nu ^2\right)\nonumber\\&
+\left.e_t^8\left(\frac{185}{48}-\frac{1073 }{288}\nu+\frac{407 }{288}\nu ^2\right)\right]\nonumber\\&
 + \left(\frac{1712}{105}+\frac{14552}{63}
e_t^2+\frac{553297}{1260} e_t^4+\frac{187357}{1260}
e_t^6+\frac{10593}{2240} e_t^8\right)\nonumber\\&
\qquad\times
\left.\left[ - C +\frac{35}{107}\pi^2 - \frac{1}{2}\ln\left(16 x\right) +
\ln\left(\frac{1+\sqrt{1-e_t^2}}{2(1-e_t^2)}\right)\right]\right\}
-\frac{116761}{3675}\kappa(e_t)\,,
\end{align}
in which again $e_t=e_t^{\rm ADM}$.
A useful internal consistency check of the algebraic correctness of
different coordinate representations of the energy flux, is the
verification that the equality of Eqs.~\eqref{AvEMha}
and~\eqref{AvEADMa} holds if and only if we have the transformation
between the time eccentricities $e_t^\mathrm{MH}$ and
$e_t^\mathrm{ADM}$ given by
\begin{align}\label{etMhtoetADM} 
\frac{e_t^\mathrm{MH}}{e_t^\mathrm{ADM}} =& 1+\frac{x^2}{1-e_t^2}
\left(-\frac{1}{4}-\frac{17}{4}\nu \right)\nonumber\\&
+\frac{x^3}{(1-e_t^2)^2}\left(-\frac{1}{2}
+\left[-\frac{16739}{1680}+\frac{21}{16}\pi^2\right]\nu +
\frac{83}{24}\nu^2+e_t^2\left(-\frac{1}{2}-\frac{249}{16}\nu
+\frac{241}{24}\nu^2\right)\right]\,.
\end{align}
(There is no ambiguity in not having a label on the $e_t$ in the 2PN
and 3PN terms above.) We find that the relation~\eqref{etMhtoetADM} is
perfectly equivalent to what is predicted from using different QK
representations of the motion, namely Eq.~\eqref{diffet} together
with \eqref{xepsj}.
%
\subsection{Gauge invariant formulation}\label{avgE-gi}
%
In the previous section, the averaged energy flux was represented using
$x$ -- a gauge invariant variable defined by~\eqref{x}
-- and the
eccentricity $e_t$ which however is coordinate dependent (but is
useful in extracting the circular limit of the result). In the present
Section we provide a gauge invariant formulation of the energy
flux.

Perhaps the most natural choice is to express the result in terms of
the conserved energy $E$ and angular momentum $J$ (per unit of reduced
mass), or, rather, in terms of the pair of rescaled variables
($\varepsilon$, $j$) defined by Eqs.~\eqref{eps}
and~\eqref{j}. However there are other possible choices for a couple
of gauge invariant quantities. As we have seen in Eqs.~\eqref{nK} the
mean motion $n$ and the periastron precession $K$ are gauge invariant
so we may define as our first choice the pair of variables ($x$,
$\iota$), where we recall that $x$ is related to the orbital frequency
$\omega=K\,n$ by Eq.~\eqref{x}, and where we define
\begin{equation}
\iota \equiv \frac{3x}{k}\,,
\end{equation}
with $k\equiv K-1$. Here we have introduced a factor 3 so that $\iota$
reduces to $j$ in first approximation (\textit{i.e.} when
$\varepsilon\rightarrow 0$). To 3PN order this parameter is related to
the energy and angular momentum variables $\varepsilon$ and $j$ by
%
\begin{eqnarray}
\label{lepsj}
\iota &=& {j +\varepsilon \left\{-\frac{27}{4}+\frac{5}{2}\nu
-j\frac{5}{12}\nu \right\}}\nonumber\\&& {+\varepsilon^2
\left\{\frac{205}{16}+\left[-\frac{1201}{48}+\frac{41}{128}\pi
^2\right]\nu +\frac{35}{24}\nu ^2+j^{\frac{1}{2}}(-5+2\nu
)+j\left(\frac{35}{16}+\frac{1}{72}\nu
^2\right)\right.}\nonumber\\&&
{\left.+\frac{1}{j}\left(-\frac{331}{16}+\left[\frac{725}{12}-\frac{205}{128}\pi^2\right]\nu +\frac{15}{8}\nu ^2\right)\right\}}\nonumber\\&&
{+\varepsilon^3
\left\{\frac{495}{64}+\left[-\frac{1145}{24}+\frac{205}{512}\pi
^2\right]\nu +\left(\frac{2341}{72}-\frac{451}{1536}\pi ^2\right)\nu
^2-\frac{415}{144}\nu ^3\right.}\nonumber\\&&
{+j^{\frac{1}{2}}\left(\frac{95}{8}-\frac{115}{24}\nu
+\frac{17}{12}\nu
^2\right)+j\left(-\frac{415}{192}-\frac{385}{192}\nu
-\frac{5}{32}\nu ^2+\frac{161}{1296}\nu ^3\right)}\nonumber\\&&
{+\frac{1}{j^{\frac{1}{2}}}\left(-\frac{5}{4}+\left[\frac{202}{9}-\frac{41}{96}\pi
^2\right]\nu
\right)+\frac{1}{j}\left(-\frac{12345}{32}+\left[\frac{147283}{192}-\frac{3895}{512}\pi
^2\right]\nu \right.}\nonumber\\&&
{\left.+\left[-\frac{77945}{288}+\frac{4715}{1536}\pi ^2\right]\nu
^2+\frac{445}{32}\nu ^3\right)}\nonumber\\&&
{\left.+\frac{1}{j^2}\left(\frac{193351}{192}+\left[-\frac{165835}{96}+\frac{7175}{256}\pi
^2\right]\nu +\left[\frac{1300}{3}-\frac{1025}{128}\pi ^2\right]\nu
^2-\frac{25}{4}\nu ^3\right)\right\}}\,.
\end{eqnarray}
We have performed two calculations of the gauge-invariant result, in
terms of the variables ($x$, $\iota$), starting from the expression of
the averaged flux in either MH and ADM coordinates. The instantaneous
part of the flux takes the form
\begin{equation}\label{AvExi}
\langle\,\mathcal{F}_\mathrm{inst}\rangle =\frac{32 c^5}{5
G}\,\nu^2\,x^5\,\iota^{-13/2}\biggl(\mathcal{I}_\mathrm{N} +
x\,\mathcal{I}_\mathrm{1PN} + x^2\,\mathcal{I}_\mathrm{2PN} +
x^3\,\mathcal{I}_\mathrm{3PN} \biggr)\,,
\end{equation}
in which the PN coefficients are polynomials of $\iota$ and the mass
ratio $\nu$ and given by
\begin{subequations}\label{AvExkpa}\begin{align}
\label{AvExkp0} \mathcal{I}_\mathrm{N} =&\frac{425}{96}\iota^3
-\frac{61}{16}\iota^4 +\frac{37}{96}\iota^5\,,\\
\label{AvExkp1} \mathcal{I}_\mathrm{1PN}  =& \left(-\frac{289}{3}
+\frac{3605}{384}\nu\right)\iota^2 +\left(\frac{1865}{24}
+\frac{3775}{384}\nu \right)\iota^3\nonumber\\&
+\left(-\frac{5297}{336} -\frac{2725}{384}\nu\right)\iota^4
+\left(\frac{139}{112} +\frac{259}{1152} \nu \right)\iota^5\,,\\
\label{AvExkp2} \mathcal{I}_\mathrm{2PN} =&
\left(\frac{267725837}{258048}
+\left[\frac{1440583}{2304}-\frac{609875}{24576} \pi ^2\right] \nu
+\frac{24395}{1024} \nu ^2\right)\iota\nonumber\\&
+\left(-\frac{51894953}{82944}
+\left[-\frac{583921}{512}+\frac{497125}{24576} \pi ^2\right] \nu
+\frac{1625}{48} \nu ^2\right)\iota^2\nonumber\\&
+\left(\frac{49183667}{387072}
+\left[\frac{14718145}{32256}-\frac{32595}{8192} \pi ^2\right] \nu
+\frac{37145}{4608} \nu ^2\right)\iota^3\nonumber\\&
+\left(-\frac{305}{16} +\frac{61}{8}\nu\right)\iota^{7/2}
+\left(-\frac{2145781}{64512}\right.\nonumber\\&
\left.+\left[-\frac{505639}{10752}+\frac{1517}{8192} \pi ^2\right] \nu
-\frac{105}{16} \nu ^2\right)\iota^4\nonumber\\& +\left(\frac{185}{48}
-\frac{37}{24}\nu\right)\iota^{9/2}+\left(\frac{744545}{258048}
+\frac{19073}{32256}\nu +\frac{2849}{27648}
\nu^2\right)\iota^{5}\,,\\
\label{AvExkp3} \mathcal{I}_\mathrm{3PN} =&
\frac{149899221067}{7741440}
+\left[-\frac{186950547065}{3096576}+\frac{46739713}{32768} \pi
^2\right] \nu \nonumber\\&
+\left[\frac{66297815}{6144}-\frac{8315825}{32768} \pi ^2\right] \nu
^2-\frac{415625}{12288} \nu^3-\frac{161249}{192}\iota^{1/2}\nonumber\\&
+\left(-\frac{66998702987}{2073600}
+\left[\frac{71728525525}{1032192}-\frac{117241181}{98304} \pi
^2\right] \nu \right.\nonumber\\&
\left.+\left[-\frac{24611099}{2304}+\frac{6633185}{49152} \pi
^2\right] \nu ^2+\frac{4346075}{12288} \nu
^3\right)\iota+\frac{3727559}{2880}\iota^{3/2}\nonumber\\&
+\left(\frac{4774135897}{322560}
+\left[-\frac{332003303819}{13934592}+\frac{29862965}{114688} \pi
^2\right] \nu \right.\nonumber\\&
\left.+\left[\frac{15103071}{7168}-\frac{3258475}{294912}\pi ^2\right]
\nu ^2-\frac{2249695}{18432} \nu ^3\right)\iota^2\nonumber\\&
+\left(-\frac{928043}{5760}+\left[-\frac{1879}{1152}-\frac{2501}{1536}
\pi ^2\right] \nu -\frac{5605}{192} \nu ^2\right)\iota^{5/2}\nonumber\\&
+\left(-\frac{2740721737}{1290240}
+\left[\frac{225135517}{73728}-\frac{8351167}{688128} \pi ^2\right]
\nu \right.\nonumber\\&
\left.+\left[\frac{6154165}{64512}-\frac{615}{1024} \pi ^2\right] \nu
^2+\frac{298895}{6144} \nu ^3\right)\iota^3\nonumber\\&
+\left(-\frac{3913177}{37800}
+\left[-\frac{351499}{12096}+\frac{1517}{4608} \pi ^2\right] \nu
+\frac{1153}{32} \nu ^2\right)\iota^{7/2}\nonumber\\&
+\left(\frac{1758850201}{141926400}
+\left[-\frac{186455099}{1032192}+\frac{68757}{229376} \pi ^2\right]
\nu \right.\nonumber\\&
\left.+\left[-\frac{5900711}{387072}-\frac{1517}{16384} \pi ^2\right]
\nu ^2-\frac{2568655}{331776} \nu ^3\right)\iota^4\nonumber\\&
+\left(\frac{51335}{2688} -\frac{10951}{2688}\nu -\frac{481}{192} \nu
^2\right)\iota^{9/2}\nonumber\\& +\left(\frac{2635805}{405504}
+\frac{891535}{3096576}\nu +\frac{4537}{27648}
\nu^2+\frac{106375}{995328} \nu ^3\right)\iota^5\nonumber\\&
+\left(\frac{161249}{192}-\frac{125939}{80}\iota
+\frac{263113}{288}\iota^2-\frac{168953}{1008} \iota^3+\frac{10593}{2240}
\iota^4\right)\ln\left[\frac{x}{x_0}\frac{1+\sqrt{\iota}}{2 \iota}\right]\,.
\end{align}
\end{subequations}
Similarly, we can also obtain the equivalent
expression of the flux in terms of the
rescaled variables ($\varepsilon$, $j$) defined by Eqs.~\eqref{eps}
and~\eqref{j}. 

The hereditary part of the flux given
by~\eqref{hered1}--\eqref{hered2} is straightforwardly added. In this 
part we have simply to replace $e_t$ by its expression in terms of
$x$ and $\iota$ at the 1PN order, namely
\begin{equation}
e_t = \Biggl[1-\iota+ x\, \biggl\{ -\frac{35}{4} +\frac{9}{2}\,\nu +
\iota \left(  \frac{17}{4}-\frac{13}{6} \,\nu \right)\biggr\} 
\Biggr]^{1/2}\,.
\end{equation}
(At this order there is no difference between MH and ADM coordinates.)
Note also that with the latter choice of gauge-invariant variables
the circular-orbit limit is not directly readable from the
expressions. However, it can be easily obtained
by 
using the expression for the variable
$j_\odot$ as reduced to circular orbits in terms of $\varepsilon$,
Eq.~\eqref{jcirc}.
\section{The test particle limit of the 3PN energy flux}\label{test}
An important check on our result is the test particle limit for which
the energy flux in the eccentric orbit case is available (to second
order in the eccentricity) from computations based on perturbation
theory around a Schwarzschild background. We compare the end result of
our computation -- composed of the instantaneous terms and the
hereditary terms computed in paper~I -- with the result obtained in
Ref.~\cite{Japanese}. Thus, we take the test particle limit of our
result (\textit{i.e.} $\nu\equiv\mu/m\rightarrow 0$), say in the form
given by Eqs.~\eqref{AvEMh}--\eqref{AvEMha} in which $e_t\equiv
e_t^\mathrm{MH}$, and expand it in powers of $e_t$ retaining only
terms up to $e_t^2$. The instantaneous contribution to the energy flux
in the test mass limit is then given by
\begin{align}
\langle\mathcal{F}_\mathrm{inst}\rangle=&\frac{32 c^5}{5
G}\,\nu^2\,x^5\left\{1-\frac{1247}{336}x-\frac{44711}{9072}x^2
+\left[\frac{1266161801}{9979200}+\frac{1712}{105}\ln\left(\frac{x}{x_0}
\right)\right]x^3\right.\\& \left.+e_t^2
\left(\frac{157}{24}-\frac{187}{168}x-\frac{84547}{756}x^2
+\left[\frac{22718275589}{9979200}+\frac{106144}{315}\ln
\left(\frac{x}{x_0}\right)\right]x^3\right)
+\mathcal{O}(\nu)\right\}+\mathcal{O}\left(e_t^4\right)\,.\nonumber
\end{align}
On the other hand, the hereditary contribution has been reported in
Eqs.~\eqref{hered1}--\eqref{hered2} and admits the test-mass limit
\begin{eqnarray}
\langle\mathcal{F}_\mathrm{hered}\rangle&=& \frac{32 c^5}{5
G}\,\nu^2\,x^5 \Biggl\{
4\pi\,x^{3/2}\,\varphi(e_t)-\frac{8191}{672}\,\pi\,x^{5/2}\,\psi(e_t)\nonumber\\
&&+x^3\left[ -\frac{116761}{3675}\,\kappa(e_t) +\left[ \frac{16}{3}
\,\pi^2 -\frac{1712}{105}\,C - \frac{1712}{105}\ln\left(\frac{4
x^{3/2}}{x_0}\right)\right] F(e_t)\right]+\mathcal{O}(\nu)\Biggr\}\,.
\end{eqnarray} 
To proceed further, all the enhancement functions should be expanded
up to power $e_t^2$. This is easy for $F(e_t)$ which is known
analytically from Eq.~\eqref{Fet} and we have
\begin{equation}\label{Fexp}
F\left(e_t\right) =
1+\frac{62}{3}\,e_t^2+\mathcal{O}\left(e_t^4\right)\,.
\end{equation}
The other enhancement functions are only known numerically for general
eccentricity. We have however succeeded in obtaining analytically their
leading correction term $e_t^2$ by implementing our calculation of the
tails in paper~I  at order $e_t^2$ from the
start. The results we thereby obtained [Eqs.~(6.8) of paper~I] are
\begin{subequations}\label{enhancexp}\begin{eqnarray}
\label{phiexp} \varphi\left(e_t\right)&=& 1+\frac{2335
}{192}\,e_t^2+\mathcal{O}\left(e_t^4\right)\,,\\
\label{psiexp} \psi\left(e_t\right)&=& 1-\frac{22988
}{8191}\,e_t^2+\mathcal{O}\left(e_t^4\right)\,,\\
\label{kappaexp} \kappa\left(e_t\right)&=& 1+\left(\frac{62}{3}
-\frac{4613840}{350283}\ln 2+\frac{24570945}{1868176}\ln 3\right)\,e_t^2
+\mathcal{O}\left(e_t^4\right)\,. 
\end{eqnarray}\end{subequations}
[We do not need $\zeta(e_t)$ here since it is in factor of a $\nu$-dependent
term.]
Our final result to $\mathcal{O}(\nu)$ and
$\mathcal{O}\left(e_t^4\right)$ is therefore
\begin{align}\label{fluxnu0}
\langle\mathcal{F}\rangle=&\frac{32 c^5}{5
G}\,\nu^2\,x^5\Biggl\{1-\frac{1247}{336}x+4\pi\,x^{3/2}-\frac{44711}{9072}\,x^2
-\frac{8191}{672}\,\pi\,x^{5/2}\nonumber\\&\quad
+\left[\frac{6643739519}{69854400}+\frac{16}{3} \,\pi^2
-\frac{1712}{105}\,C - \frac{856}{105}\ln\left(16 x\right)\right]\,x^3
\nonumber\\+& e_t^2\Biggl(\frac{157}{24}
-\frac{187}{168}\,x+\frac{2335}{48}\,\pi\,x^{3/2}-\frac{84547}{756}\,x^2
+\frac{821}{24}\,\pi\,x^{5/2}\nonumber\\&\quad+\left[\frac{113160471971}{69854400}
+\frac{18832}{45}\ln 2-\frac{234009}{560}\ln 3
\right.\nonumber\\&\qquad\left.+\frac{992}{9} \,\pi^2
-\frac{106144}{315}\,C - \frac{53072}{315}\ln\left(16
x\right)\right]\,x^3 \Biggr) +\mathcal{O}\left(e_t^4\right) +
\mathcal{O}\left(\nu\right) \Biggr\}\,.
\end{align}
The above expression is in terms of our chosen eccentricity $e_t$. One
should note that the ``Schwarzschild'' eccentricity $e$ appearing in
the black-hole perturbation theory~\cite{Japanese} is \textit{a
priori} different from $e_t$; therefore the above result can only be
compared modulo a transformation of these eccentricities. We find that
indeed Eq.~\eqref{fluxnu0} is equivalent to the black-hole
perturbation result given by Eq.~(180) of~\cite{Japanese}, if and only
if the two eccentricities are linked together by
\begin{equation}\label{ete} e_t^2=
e^2\left(1 - 6 x + 4 x^2 - 8 x^3\right)\,.
\end{equation}
(Recall that $e_t=e_t^\mathrm{MH}$ here.)
\section{Concluding remarks}
The instantaneous contributions to the 3PN gravitational wave
luminosity from the inspiral phase of a binary system of compact
objects moving in an elliptical orbit is computed using the Multipolar
post-Minkowskian wave generation formalism\footnote{The instantaneous part of the 3PN gravitational wave flux of angular momentum and linear momentum
 from inspiralling
compact binaries moving on elliptical orbits has been computed~\cite{ArunThesis,ABIQ07AMF}.}.The non-trivial inputs for
this calculation include the mass octupole and current quadrupole at
2PN order for general orbits and the 3PN accurate mass
quadrupole. Using the 3PN quasi-Keplerian representation of elliptical
orbits obtained recently the flux is averaged over the binary's
orbit. The instantaneous part of the energy flux is computed in the
standard harmonic coordinate system (with logarithms). For technical
reasons the average over an orbit of the instantaneous contributions
is presented in other coordinate systems: Modified harmonic
coordinates (without logarithms) and ADM coordinates. Alternative
\textit{gauge invariant} expressions are also provided. Supplementing
the instantaneous contributions of this paper by the important
hereditary contributions arising from tails, tails-of-tails and tails
squared terms calculated in paper~I~\cite{ABIQ07Tail}, the complete
energy flux has been obtained.

For binaries moving on circular orbits the 3PN energy flux agrees with
that computed in~\cite{BIJ02}. However the circular-orbit results are
known to the higher 3.5PN order~\cite{BIJ02}. The extension of the
3.5PN term to eccentric orbits would be interesting, but some
uncomputed modules remain in the general formalism to compute the multipole
moments for general sources
 required for the 3.5PN generation in the eccentric orbit
case. We leave this to a future investigation.

\acknowledgments L.B. and B.R.I. thank the Indo-French Collaboration
(IFCPAR) under which this work has been carried out. M.S.S.Q acknowledges the
Indo-Yemen cultural exchange programme. 
B.R.I. acknowledges the
hospitality of the Institut Henri Poincar\'e and Institut des Hautes
Etudes Scientifiques during the final stages of the writing of the
paper.
Almost all algebraic
calculations leading to the results of this paper are done with the
software MATHEMATICA.


\bibliography{thi-minimal}
\end{document}